\documentclass[%
 reprint,
superscriptaddress,
 amsmath,amssymb,
 aps,
floatfix,
prd
]{revtex4-1}

\usepackage{graphicx}% Include figure files
\usepackage{dcolumn}% Align table columns on decimal point
\usepackage{bm}% bold math
\usepackage{appendix}
\usepackage{xcolor}
\usepackage{ulem}

\begin{document}

%\preprint{APS/123-QED}

\title[Light mediator]{Massive neutrino self-interactions with a light mediator in cosmology}

   \author{Jorge Venzor}\email{jorge.venzor@cinvestav.mx}
    \affiliation{ Departamento de F\'{\i}sica, Centro de Investigaci\'on y de Estudios Avanzados del I.P.N. \\
    Apartado Postal 14-740, 07000, Ciudad de M\'exico, M\'exico. }
    \author{Gabriela Garcia-Arroyo}\email{0543074c@umich.mx}
    \affiliation{Dpto. Ingenier\'{\i}a Civil, Divisi\'on de Ingenier\'{\i}a, Universidad de Guanajuato, Gto., CP 36000, M\'exico.}
    \author{Abdel P\'erez-Lorenzana}\email{aplorenz@fis.cinvestav.mx}
    \affiliation{ Departamento de F\'{\i}sica, Centro de Investigaci\'on y de Estudios Avanzados del I.P.N. \\
    Apartado Postal 14-740, 07000, Ciudad de M\'exico, M\'exico. }
    \author{Josue De-Santiago}\email{Josue.desantiago@cinvestav.mx}
    \affiliation{ Departamento de F\'{\i}sica, Centro de Investigaci\'on y de Estudios Avanzados del I.P.N. \\
    Apartado Postal 14-740, 07000, Ciudad de M\'exico, M\'exico. }
    \affiliation{Consejo Nacional de Ciencia y Tecnolog\'{\i}a.
    Av. Insurgentes Sur 1582, 03940, Ciudad de M\'exico, M\'exico}

\date{\today}

\begin{abstract}
Nonstandard self-interactions can alter the evolution of cosmological neutrinos, mainly by damping free streaming, 
which should leave traces in cosmological observables. Although overall effects are opposite to those produced by neutrino mass and a larger $N_{\rm eff}$, they cannot be totally canceled by these last. 
We harness cosmological data that includes Cosmic Microwave Background from Planck 2018, BAO measurements, local $H_0$, Ly-$\alpha$ and SNIa, to constrain massive neutrino self-interactions with a very light scalar mediator.
We find that the effective coupling constant, at the 95\% C.L., should be $g_{\rm eff}< 1.94 \times 10^{-7}$ for only Planck 2018 data and $1.97\times10^{-7}$ when Planck + BAO are considered. This bound  relaxes to $2.27\times 10^{-7}$ ($2.3\times 10^{-7}$) for $H_0$ ($H_0$+SNe+Ly-$\alpha$) data.
Using the Planck + BAO dataset, the $H_0$ tension lowers from 4.3$\sigma$ (for $\Lambda$CDM) to 3.2$\sigma$. The Akaike Information Criterion penalizes the self-interacting model due to its larger parameter space for Planck or Planck + BAO data, but favors the interacting model when we use local $H_0$ measurements.
A somewhat larger value for $H_0$ is preferred when we include the whole data pool, which comes accompanied with a larger value of $N_{\rm eff}$  and a more constricted bound on $\Sigma m_\nu$. 
\end{abstract}

\maketitle

%\tableofcontents
%%%%%%%%%%%%%%%%%%%%%%%%%%%%%%%%%%%%%%%%%%%%%%%%%%%%%%%%%%%%%%%
\section{\label{sec:introduction} Introduction}
%%%%%%%%%%%%%%%%%%%%%%%%%%%%%%%%%%%%%%%%%%%%%%%%%%%%%%%%%%%%%%%

Cosmology provides frontier conditions to uniquely study neutrinos.
The cosmological constraints on the standard neutrinos are consistent with three active neutrinos \cite{Bashinsky2004,Follin2015,baumann2019first},
which is consistent with the theoretical prediction $N_{\rm eff \ std}=3.045$ \cite{Mangano2005,deSalas2016,Escudero2020,Akita2020,froustey2020neff}.
Furthermore, the cosmological bound on the sum of the neutrino masses is the more restrictive so far with a consensus value of $\sum m_{\nu}< 0.12$ eV \cite{Aghanim2018Planck,Guisarma2016}.
Cosmology is also expected to robustly show a preference for the correct neutrino mass hierarchy, wherein the normal (inverted) hierarchy $m_3>m_1$ ($m_3<m_1$)~.
Once the bound on the sum of the neutrino masses reaches a precision below the minimum bound in the IH, $\sum m_{\nu}^{\rm IH}>0.0986 \pm 0.00085$ eV \cite{Loureiro2019}, a preference for one the hierarchies would be plausible \cite{Vagnozzi2017,Brinckmann2019,DiValentino2021}.

However,  neutrino cosmological constraints may change significantly in the presence of neutrino nonstandard interactions (NSI).
For instance, in such a scenario, an extra degeneracy arises with other cosmological parameters.
Additionally, the impact on the cosmological observables does depend heavily on the nature of the NSI and on the mediator mass.

A neutrino self-interaction mediated by a heavy particle is an interesting model that could ease the $H_0$ tension \cite{Bell2006,Archidiacono2014,CyrRacine2014,Lancaster2017,Kreisch2020,Park2019,Choudhury2021}. 
Late and early observations of the Hubble parameter today are, on average, not consistent with each other \cite{Verde2019,Aghanim2018Planck,ACT020,DES2018,Riess2019ApJ,H0licow2020}.
This tension is now above 4$\sigma$. If it were confirmed, new physics in the early Universe might be needed to explain the discrepancy.
In this sense, among several ideas for physics beyond the Standard Model, neutrino NSIs are particularly appealing because their validity can soon be proved right or wrong by experiments and astrophysical observations \cite{Heurtier2017,Huang2018,Blinov2019,Brune2019,brinckmann2020self,mazumdar2020flavour,Das2021,Schoneberg2019,Huang2021,Brdar2020,Esteban2021_probing}.

On the other hand, a neutrino NSI mediated by a light scalar particle is one of the preferred proposals since it could be related to the solution to the anomaly observed by the MiniBooNE and LSND experiments \cite{LSND2001,MiniBoone2018,Palomares2005,deGouvea2020}.
Although the interpretation of the excess signal produced by single electron neutrinos was recently disfavored by the MicroBooNE collaboration \cite{abratenko2021search}.
The origin of the neutrino excess observed by MiniBooNE/LSND is still to be explained and a neutrino NSI mediated by a light particle is a possible explanation.

Aside from possible solutions to the mentioned tensions and anomalies, the phenomenological consequences of neutrino NSI with a light mediator are vast.
For instance, the neutrino mass bounds may relax or completely vanish in the presence of neutrino decays, annihilations or long-range interactions via a NSI neutrino-scalar interactions \cite{Beacom2004,Hannestad2005,Chacko2020,Escudero2020relaxing,Esteban2021,Barenboim_2021,abellan2021improved}.
Furthermore, non-desirable NSI effects change $N_{\rm eff}$, by increasing \cite{Huang2018}, or reducing its value \cite{Venzor2021}.
The neutrino self-interactions effects on linear perturbation cosmology have been studied in \cite{Forastieri2015,Forastieri2019}, where the authors found a 2$\sigma$ bound for the effective neutrino-scalar coupling $g_{\rm eff}\lesssim 2\times 10^{-7}$.
So far, self-interactions with light mediators have not shown any preference for a larger $H_0$-value~\cite{Forastieri2019}.

In this paper, we study the effect of massive neutrino self-interactions mediated by a light ($m_{\phi}\lesssim 10^{-3}$ eV), or massless scalar particle on the evolution of cosmological density perturbations and temperature and matter power spectra.
We extend the previous works in Refs. \cite{Forastieri2015,Forastieri2019},
where they studied self-interactions without taking into account the neutrino mass. We compare our results to the latest cosmological observations in order to constrain the parameters of the model.
Finally, we use model comparison criteria to test the self-interacting neutrinos versus the standard cosmological model.

The rest of the paper is organized as follows, in section \ref{sec:perturb}, we introduce the computation of the neutrino self-interaction collision term, then we use the relaxation time approximation (RTA) to write down the Boltzmann hierarchy of neutrino perturbations. In section \ref{sec:constraint}, we constrain the parameter space for neutrino self-interactions using different cosmological data sets.
We conclude and highlight some future directions in section \ref{sec:conclusions}.

%%%%%%%%%%%%%%%%%%%%%%%%%%%%%%%%%%%%%%%%%%%%%%%%%%%%%%%%%%%%%
\section{\label{sec:perturb} Cosmological perturbations in the presence of neutrino self-interactions }
%%%%%%%%%%%%%%%%%%%%%%%%%%%%%%%%%%%%%%%%%%%%%%%%%%%%%%%%%%%%%

Along with the present discussion, we shall focus on the thermal history of neutrinos, its impact on the evolution of the other species, and the imprints that this last should leave on current cosmological data,  which can be 
appropriately described by the evolution of its full distribution function in phase space, through Boltzmann equation formalism.
In the standard approach, neutrinos are only subject to electroweak interactions, which keep neutrinos in thermal equilibrium well down to about $1\sim$MeV, its decoupling temperature. Once decoupled, the ultrarelativistic neutrinos free-stream, damping neutrino density fluctuations at scales below the free-streaming length, feeding $\ell\gg 2$ moments in the neutrino Boltzmann hierarchy. As a consequence, metric perturbations also get reduced as a back reaction within those scales. As they travel through matter they gravitationally pull wavefronts, phase shifting CMB power spectra towards larger scales, also damping its amplitude~\cite{Bashinsky2004,Follin2015,baumann2019first}. The signature of this effect appears as a photon sound horizon at last scattering that is slightly larger than what it would be in the absence of this effect.
On the larger scales, on the other hand, free streaming does not affect neutrino perturbations, and thus at late times, they  would evolve just as cold dark matter once they become non relativistic~\cite{Choi_2018,Follin2015,Baumann_2017}.

However, neutrino NSIs alter this picture.
A moderate neutrino NSI that remains effectively active well down to the epoch of photon decoupling and beyond would have sensible effects on the evolution of the cosmic microwave background (CMB) and matter perturbations. The final scenario does depend, however, on the intermediary mass, as previous studies had pointed out~\cite{Oldengott2015,Oldengott2017,Kreisch2020}. For a massive mediator particle the NSI competes with standard weak interactions, so that if neutrinos remain coupled to matter after weak interactions had decoupled, neutrino decoupling temperature would be delayed, as well as the free streaming length diminished. If this interaction keeps neutrinos coupled to matter long enough, it will prompt neutrinos to isotropize, which means damping of high Boltzmann moments, increasing the effects of the lower monopole and dipole moments in acoustic oscillations. Of course, in this case,
Neutrino NSI should eventually go out of equilibrium, allowing the neutrinos to free stream, unchanging the behavior on the large scale perturbations. The effect of having a massive (sterile) neutrino experimenting this class of NSI interactions was considered in Ref.~\cite{Kreisch2020}.

On the other hand, cosmological effects of lighter mediators of neutrino NSI had been considered in Refs.~\cite{Basboll2009,Oldengott2015,Forastieri2015,Forastieri2019}.
In this case, the NSI interaction rate may remain subleading compared to the expansion of the Universe at early times, but, it eventually can overtake the expansion bringing neutrinos into a late self-coupled era, which should leave an imprint on the matter and CMB spectra. Late recoupling of neutrinos among themselves may alter the distinctive phase shift and amplitude damping of CMB acoustic oscillations, on the affected scales. In addition to that, although neutrino tiny mass barely affects neutrino free streaming, neutrino abundance is proportional to it, and thus, increasing the neutrino mass amplifies mentioned free streaming effects.
Famously, the neutrino mass suppresses the matter power spectrum for small scales~\cite{Lesgourgues:2006nd}.
From this point of view, neutrino mass could compensate for a portion of the NSI effects.
However, neutrino masses had not been so far considered as part of the variables involved in the outcomes of NSI effects. This is one of the issues we would like to address along with this paper.

In this work we assume that the geometry of the space--time is described by a perturbed FLRW metric,
which in the notation by Ma and Bertschinger (MB) \cite{Ma:1995ey} and in a synchronous gauge is written as
\begin{equation}
    ds^2=a^2(\tau)\left[-d\tau^2 + (\delta_{ij}+h_{ij})dx^idx^j\right]\, ,
    \label{metric}
\end{equation}
where $a$ is the scale factor, $\tau$ the conformal time and the scalar part of the metric perturbation, $h_{ij}(x,\tau)$, is described in the Fourier space through two fields, $h=h(\vec{k},\tau)$ and $\eta=\eta(\vec{k},\tau)$,
\begin{equation}
    h_{ij}(\vec{x}, \tau)= \int d^3k e^{i\vec{k}\cdot\vec{x}} \left\{ k_ik_jh+(k_ik_j-\frac{\delta_{ij}}{3})6 \eta\right\}.
\end{equation}
%Einstein eqs?
In such a  Universe the metric potentials are coupled to the matter constituents through the Einstein equations, from which the conservation of the energy-momentum tensor is derived and, in turn, do so the continuity and Euler equations that dictate the evolution of matter.

To proceed with our analysis, we shall first elaborate on neutrino Boltzmann hierarchy equations.
For this, we start by  writing at linear order in perturbation theory the phase space distribution function, $f(x^i, q_j, \tau)$, as 
\begin{equation}\label{eq:f_def}
    f(x^i, q_j, \tau)=f_0(q)\left[1 + \Psi (x^i, q, \hat{n}_j, \tau)\right]\, ,
\end{equation}
where $x^i$ denotes the spatial coordinates,   $q_j=q \hat{n}_j$ is the comoving tri--momentum oriented along the unitary vector $\hat{n}_j$, and with magnitude $q =|{\vec{q}}|$. This last term is related to the proper momentum $\vec{p}$, through the expansion parameter such that $\vec{q} = a\vec{p}$. On the above expression,  $f_0(q)$ is the  distribution function at the background (considered in thermal equilibrium) that in the case of neutrinos is the Fermi--Dirac distribution, 
 \begin{equation}\label{eq:fermi-dirac}
  f_{0}(q, \tau)= \frac{1}{e^{\epsilon/aT}+1}  \, .
\end{equation} 
Here $T$ is the neutrino temperature and $\epsilon=\epsilon(q,\tau)=\sqrt{q^2+a^2m^2}$  the comoving energy, related to the proper energy by $\epsilon = aE$. Since the distribution function $f$ provides the number of particles in a differential volume in phase space, 
$dN = fdx^1dx^2dx^3dP^1dP^2dP^3$, with $P^i$ the canonical conjugate momentum which in the synchronous gauge is given as 
$P_i=(\delta{ij} + \frac{1}{2}h_{ij})q ^j$, in the presence of local collisions, it evolves according to the Boltzmann equation
\begin{equation}
    \frac{Df}{d\tau}=\frac{\partial f}{\partial\tau} + \frac{\partial \vec{x}}{d\tau}\cdot\frac{\partial f}{\partial \vec{x}} + \frac{\partial \vec{P}}{d\tau}\cdot\frac{\partial f}{\partial\vec{P}}  =C[f]~,
\end{equation}
where $C[f]=\left(\frac{\partial f}{\partial \tau}\right)_C$ is the collision term.

To first order,  the Boltzmann equation for the distribution \eqref{eq:f_def} in Fourier space, has the generic expression~\cite{Ma:1995ey},
\begin{equation}\label{vlasov}
    \frac{\partial \Psi}{\partial \tau} + i \frac{q}{\epsilon}(\vec{k}\cdot \hat{n})\Psi + \frac{d \ln{f_0}}{d\ln{q}}\left[\dot{\eta}-\frac{\dot{h}+6\dot{\eta}}{2}(\vec{k}\cdot \hat{n})^2\right]=\frac{1}{f_0}C[f] .%\left(\frac{\partial f}{\partial \tau}\right)_C~.
\end{equation}
Of course, for a vanishing $C[f]$  we recover the collisionless Boltzmann equation or Vlasov equation that describes free streaming neutrinos. As it is usual, to partially reduce the dimensionality of the problem we should expand the perturbation $\Psi$ into a Legendre series,
\begin{equation}
    \Psi(\vec{k},\hat{n},q,\tau) =\sum_{\ell=0}^{\infty}(-i)^{\ell}(2\ell + 1)\Psi_{\ell}(\vec{k},q,\tau)P_{\ell}(\hat{k}\cdot \hat{n})\, ,
\end{equation} 
which transforms the former equation (\ref{vlasov}) into an infinite moment hierarchy for the $\Psi_{\ell}(\vec{k},q,\tau)$ weights.

In principle, one should solve the collisional Boltzmann equation by calculating the integrals of the collision terms for the specific interaction, however as a first approach we use the relaxation time approximation (RTA) \cite{Hannestad:2000gt} in which the collision term $C[f]/f_0$ is well approximated by $-\Psi/\tau_c$, where $\tau_c$ is the mean time between collisions, which can be written as $\tau_{c}^{-1}=a\Gamma=an\langle \sigma v\rangle$. Using this approach one obtains the following Boltzmann hierarchy,
\begin{eqnarray}
\dot{\Psi}_0 &=& -\frac{qk}{\epsilon}\Psi_1+ \frac{\dot{h}}{6}\frac{d \ln f_0}{d \ln q} \, , \\
\dot{\Psi}_1 &=& \frac{qk}{3\epsilon}(\Psi_0 -2\Psi_2) \, ,\\
\dot{\Psi}_2 &=&\frac{qk}{5\epsilon}(2\Psi_1-3\Psi_3) -\left(\frac{\dot{h}}{15}+\frac{2\dot{\eta}}{5}\right)\frac{d \ln f_0}{d \ln q}-a\Gamma \Psi_2\, , \nonumber \\
&&\\
\dot{\Psi}_{l\geq 3} &=& \frac{q k}{(2l+1)\epsilon}\left(l\Psi_{l-1}-(l+1)\Psi_{l+1}\right) -a \Gamma \Psi_l\, .
\end{eqnarray}
The first two hierarchy equations, for the monopole and dipole modes, are related to  density and energy momentum conservation, which require that 
$$\int d^3q\, C[f] =0=\int d^3q \,\hat{k}\cdot \hat{n} C[f] $$
respectively, and therefore they should get no contribution from the interaction~\cite{Oldengott2015,Forastieri2015}
whose impact becomes more relevant for the higher modes, since it enters as a damping term. We adopt a phenomenological recipe for the neutrino scattering rate for a light mediator~\cite{Forastieri2019} for which we take
\begin{equation}\label{eq:Gamma_scat}
\Gamma= 0.183 ~g_{\rm{eff}}^{4}T_{\nu}\, ,
\end{equation}
with $g_{\rm{eff}}^{4}$ an effective coupling constant that generically encodes the subtleties of the interaction.

\begin{figure}
    \centering
    \includegraphics[width=0.48\textwidth]{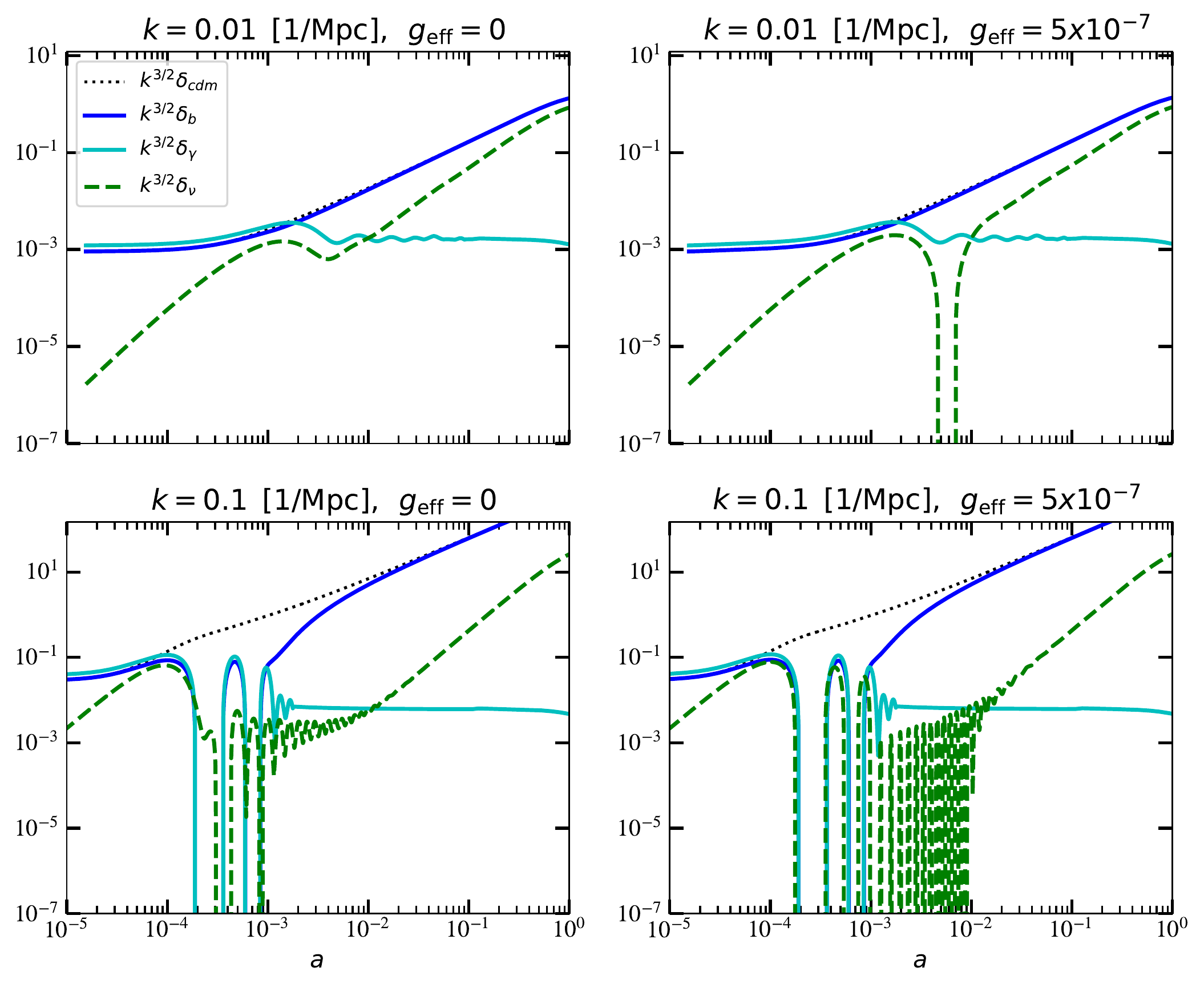}% Here is how to import EPS art
    \caption{\label{fig:deltas} Evolution of the density perturbations as a function of the scale factor in the synchronous gauge, with fixed values $\sum m_{\nu}=0.23$ eV and $N_{\rm eff}=3.046$. The value of the $k$-modes are 0.01 (top) and 0.1 1/Mpc (bottom). The interaction is turned on only in the right hand panels with $g_{\rm eff}=5\times 10^{-7}$.  }
\end{figure}

\begin{figure}
    \centering
    \includegraphics[width=0.48\textwidth]{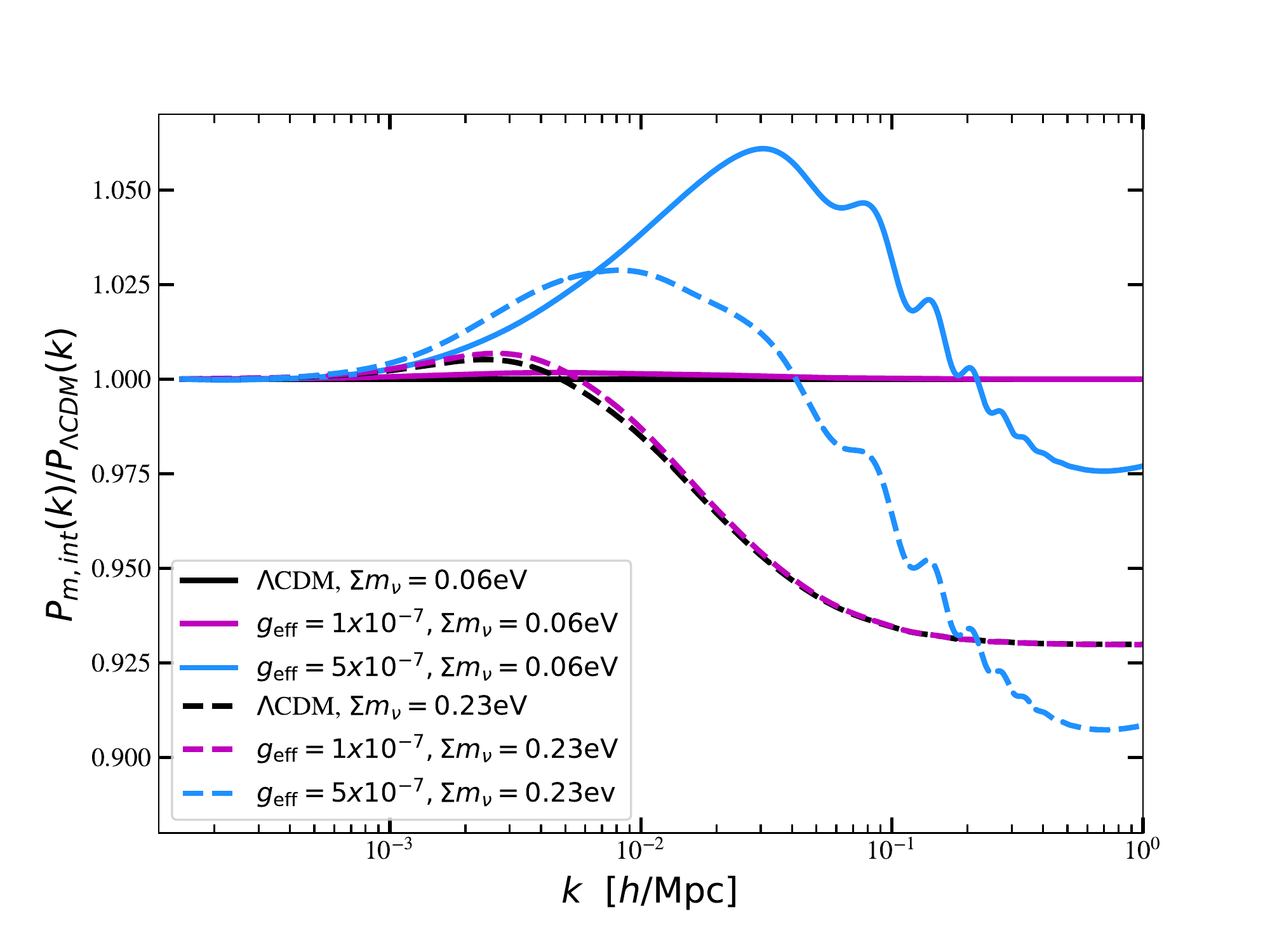}% Here is how to import EPS art
    \caption{\label{fig:mps} Effects of $g_{\rm eff}$ and $\sum m_{\nu}$ on the matter power spectrum. Solid lines correspond to $\sum m_{\nu}=0.06$ eV, and dashed ones correspond to $\sum m_{\nu}=0.23$ eV. Black indicates non-interacting neutrinos, while blue and purple lines are for self-interacting neutrinos.}
\end{figure}

Once this formalism has been introduced, it is possible to obtain the numerical solutions of the density contrasts and understand the physical effect of the interaction term.
In order to explore the self-interacting neutrino effects, we first show the effect on the density contrasts of the standard cosmological species.
In Fig. \ref{fig:deltas} we show cold dark matter, photon, baryonic and neutrino perturbations at different $k$ modes and for two $g_{\rm{eff}}$ values, 0 and $5 \times 10^{-7}$. For $g_{\rm{eff}}=0$, there is no interaction and the contrast of massive neutrinos behaves as expected (left panels), however, when the interaction turns on (right panel) the neutrino fluid undergoes acoustic oscillations because free streaming is not efficient enough to damp the perturbations.
This lack of free-streaming ultimately will translate into non-desired changes that could be partially compensated by changing other neutrino or cosmological parameters.

Changes in density perturbations are carried over the matter power spectrum, for example, in Fig. \ref{fig:mps} we show that a larger sum of neutrino masses produce an attenuation at small scales (as expected).
However, the value of the coupling constant plays an opposite role since larger values of $g_{\rm eff}$ are associated with an increasing power spectrum, so that, bigger couplings could be observationally valid if neutrino masses are increased too.
As a matter of fact, this non-trivial effect, regarding its $k$-dependence, was also observed in the heavy mediator scenario \cite{Kreisch2020}.
As we can see in the figure, the neutrino mass and the coupling cannot vanish each other effect, thus, we expect cosmology to prefer negligible values of these two parameters when employing data to constrain them.

\begin{figure}
    \centering
    \includegraphics[width=0.48\textwidth]{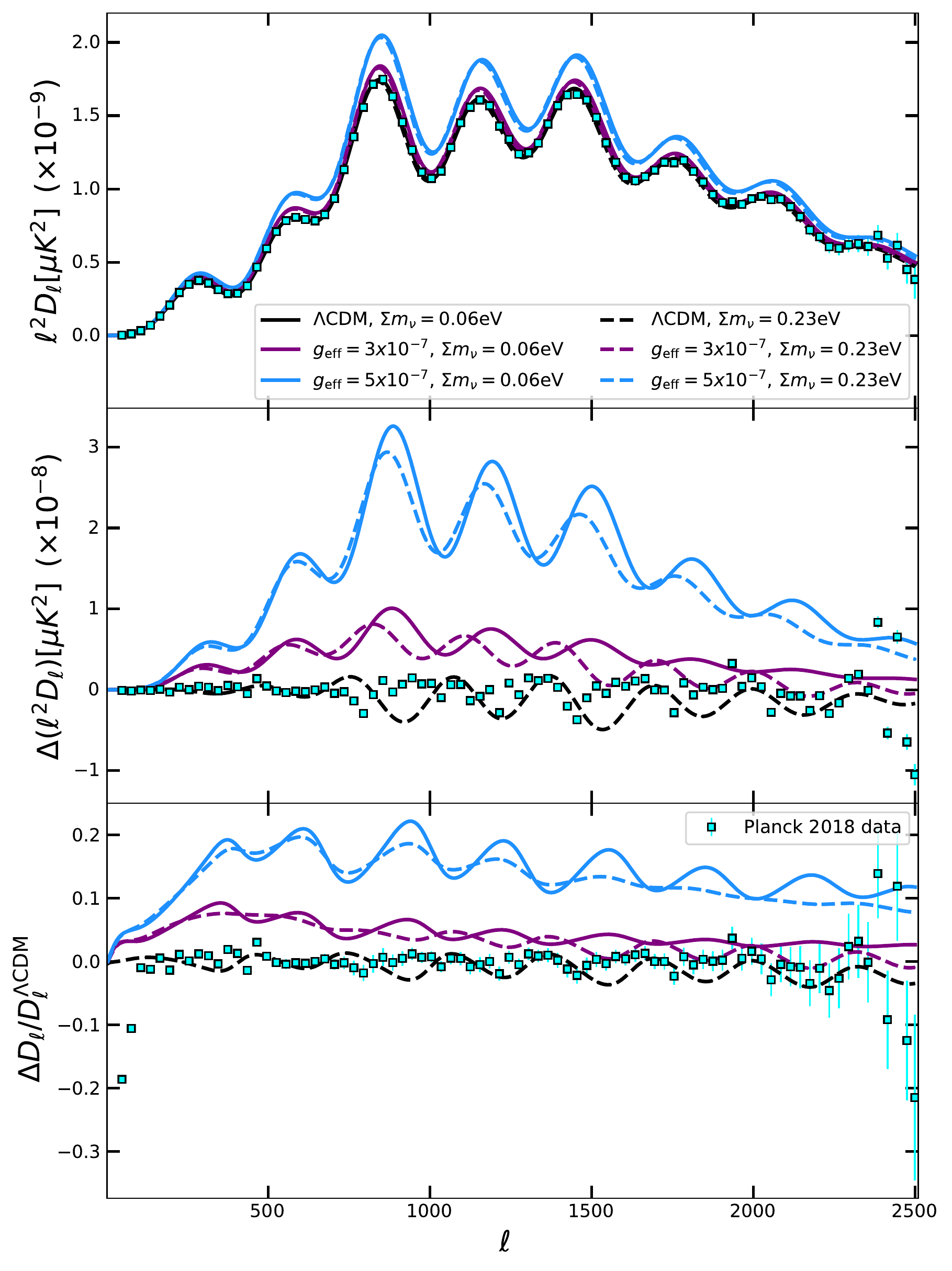} % Here is how to import EPS art
   \caption{\label{fig:d_ell}Temperature power spectrum of self-interacting massive neutrinos. Solid lines correspond to $\sum m_{\nu}=0.06$ eV, while dashed lines indicate $\sum m_{\nu}=0.23$ eV. Black indicates non interacting neutrinos, while colored lines are for self-interacting neutrinos. In the middle and bottom panels $\Delta D_{\ell}$ means $D_{\ell, \rm{NSI}}-D_{\ell, \rm{\Lambda CDM}}$.}
\end{figure}

\begin{figure}
    \centering
    \includegraphics[width=0.48\textwidth]{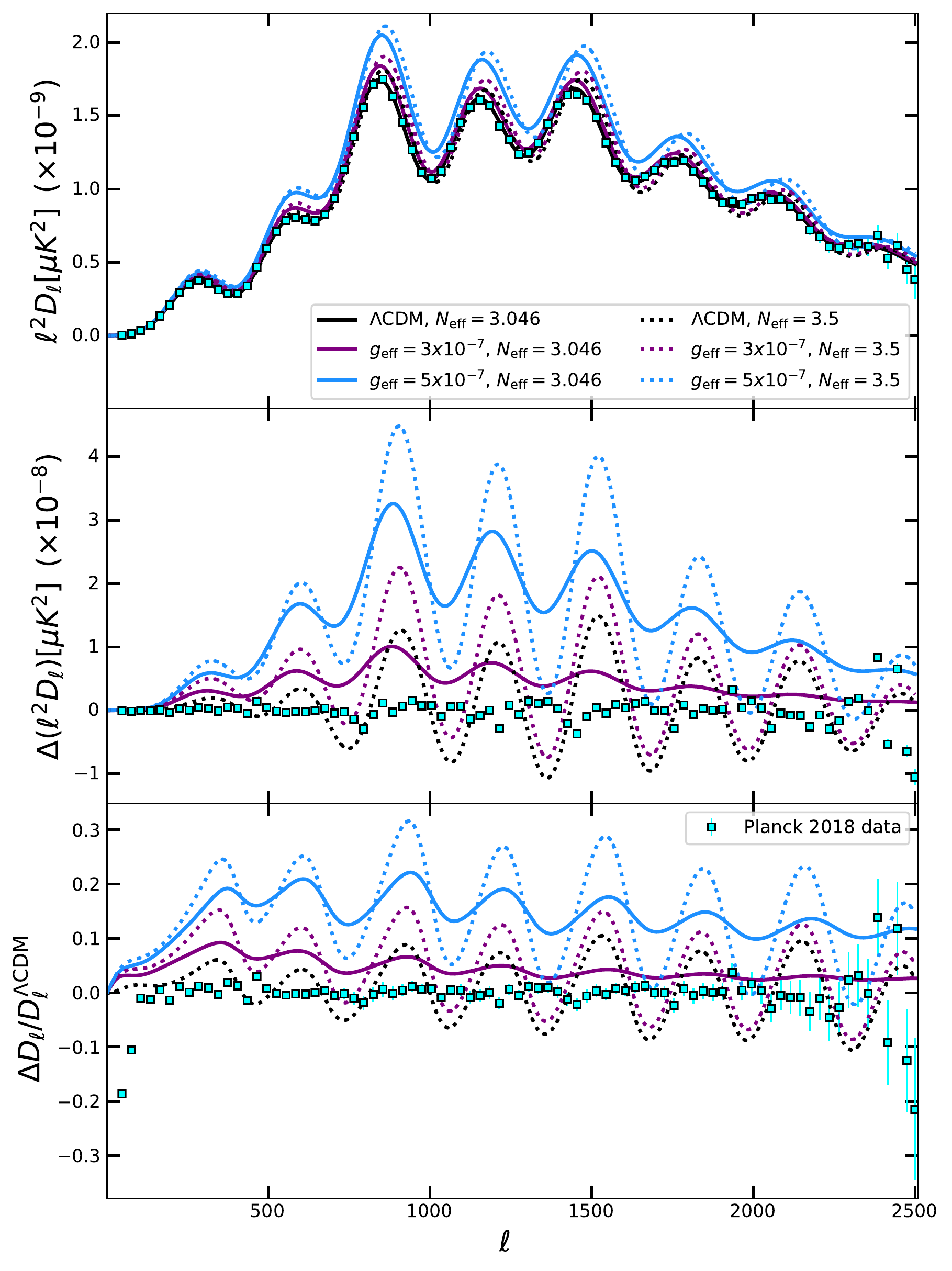}% Here is how to import EPS art
    \caption{\label{fig:d_neff}Temperature power spectrum of self-interacting massive neutrinos. Solid lines correspond to $N_{\rm eff}=3.046$, while dotted lines indicate $N_{\rm eff}=3.5$. Black indicates non-interacting neutrinos, while colored lines are for self-interacting neutrinos. }
\end{figure}

\begin{figure}
    \centering
    \includegraphics[width=0.48\textwidth]{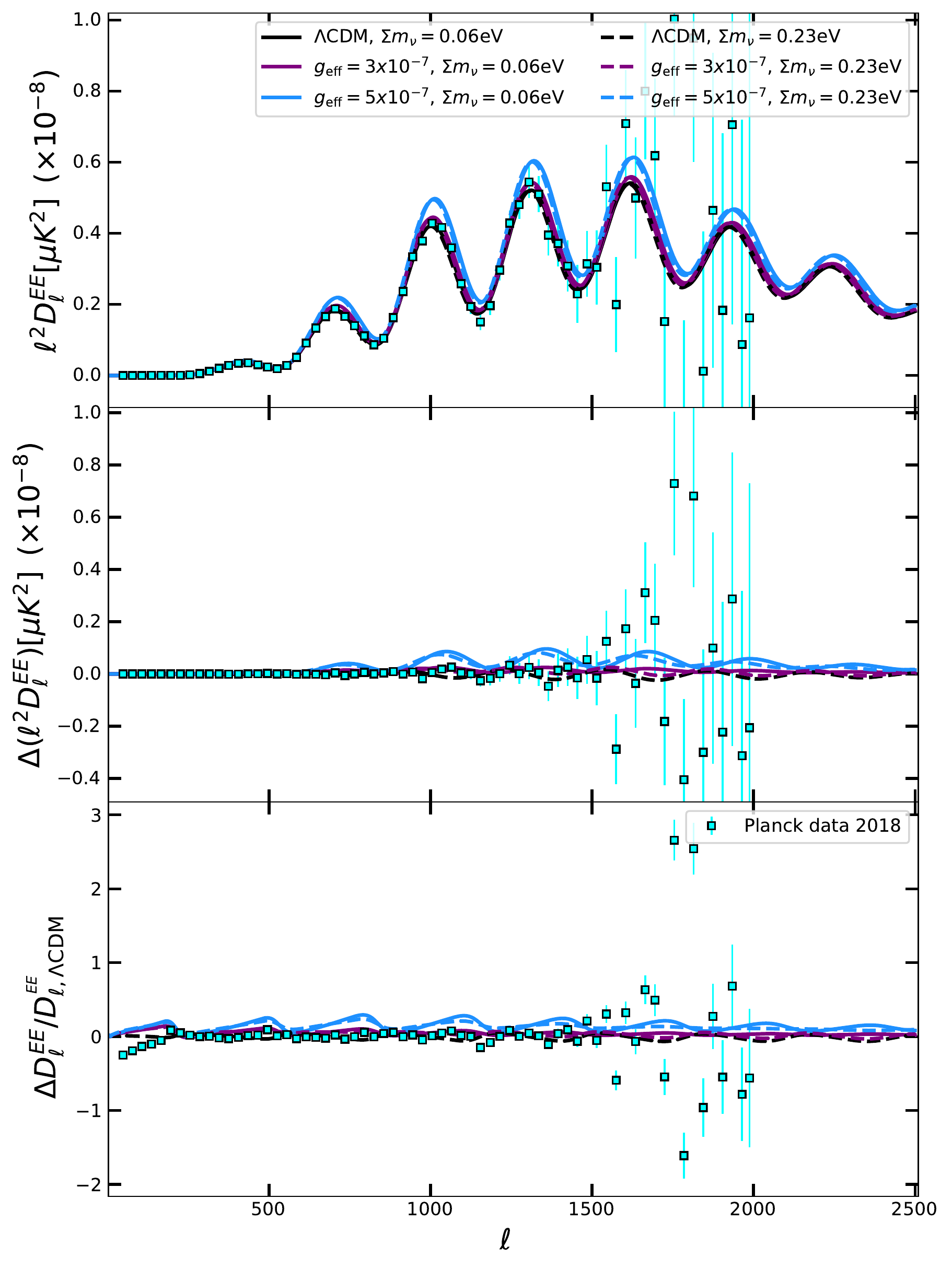}% Here is how to import EPS art
    \caption{\label{fig:d_ell_ee}EE polarization spectra of self-interacting massive neutrinos. Solid lines correspond to $\sum m_{\nu}=0.06$ eV, while dashed lines indicate $\sum m_{\nu}=0.23$ eV. Black indicates non-interacting neutrinos, and colored lines are for self-interacting neutrinos. }
\end{figure}

Massive self-interacting neutrinos  through eq.\eqref{eq:Gamma_scat} also imprint an effect in the CMB spectra, as shown in Figs.(\ref{fig:d_ell}-\ref{fig:d_ell_ee}).
In Figs. \ref{fig:d_ell} and \ref{fig:d_neff} we show the TT spectrum for different $g_{\rm eff}$-values, while also changing values of $\sum m_{\nu}$ and $N_{\rm eff}$ respectively.
We notice that $g_{\rm eff}$ enhances the spectrum for high multipoles, while the neutrino mass damps and shifts the overall spectrum to the left. Furthermore, $N_{\rm eff}$ causes a phase shift in the acoustic peaks.
In Fig. \ref{fig:d_ell_ee} we also show that the neutrino self-interaction enhances the polarization spectra, however, notice that for the selected values it is harder to distinguish from $\Lambda$CDM.
As in the MPS graph, we do not observe any non-null combination of parameters that could mimic $\Lambda$CDM in the CMB, this ultimately will reflect on different model-fit results that can be distinguished from the standard cosmological model. 

In the next section, we are going to show the results on the global fits for the self-interacting neutrino model while allowing all three neutrino parameters ($g_{\rm eff}$, $N_{\rm eff}$, $\sum m_{\nu}$) to be free.


%%%%%%%%%%%%%%%%%%%%%%%%%%%%%%%%%%%%%%%%%%%%%%%%%%%%%%%%%%%%%%%%%%%%%%%%%%%%%%%%%%
\section{\label{sec:constraint} Parameter constraints and degeneracies}
%%%%%%%%%%%%%%%%%%%%%%%%%%%%%%%%%%%%%%%%%%%%%%%%%%%%%%%%%%%%%%%%%%%%%%%%%%%%%%%%%%

In this section, we use different sets of cosmological observations to constrain the parameters of the NSI scenario. The observations include CMB power spectra from Planck 2018, which include the combined TT, TE, EE, low E, and lensing likelihoods \cite{Planck:2018nkj}.
Baryonic acoustic oscillations (BAO) measurements from BOSS DR12 ---both from galaxy \cite{BOSS:2016wmc} and Lyman-$\alpha$ forest correlations \cite{Bautista:2017wwp,duMasdesBourboux:2017mrl}---, 6DF \cite{Beutler:2011hx} and MGS \cite{Ross:2014qpa}. In the following we refer to the galaxy measurements simply as BAO and to the BAO Lyman-$\alpha$ measurements simply as Ly-$\alpha$. Local $H_0$ measurements from SH0ES \cite{Riess2019ApJ} are also included as well as SNIa data from Pantheon \cite{Pan-STARRS1:2017jku}.

\begin{figure*}
\includegraphics[width=0.7\textwidth]{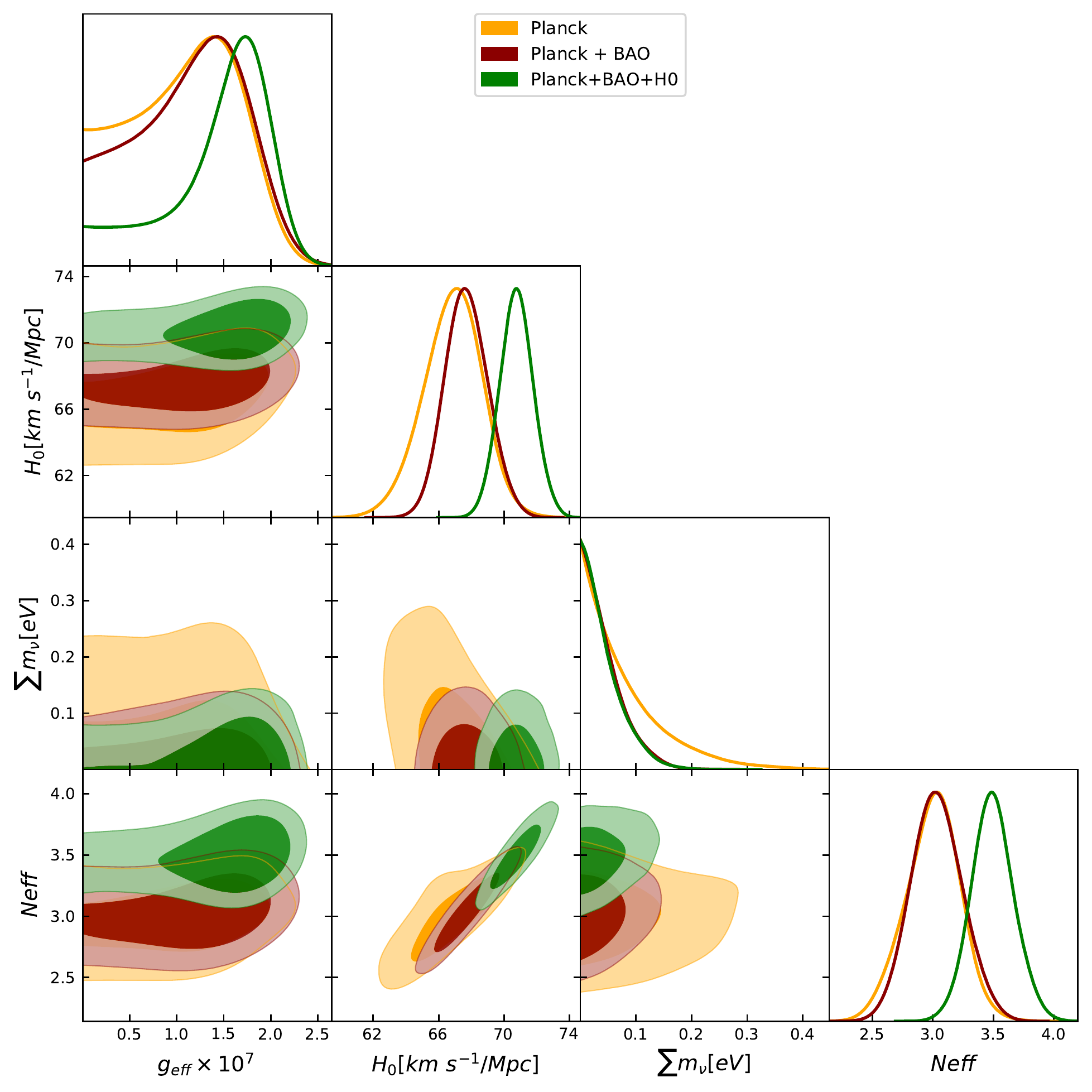}% Here is how to import EPS art
\caption{\label{fig:triangle}  Observational constraints for the NSI model, there are included the main parameters at 68\% and 95\% confidence limits; using Planck, Planck+BAO and, Planck+BAO+$H_0$ data combinations. We see that the inclusion of local $H_0$ data favors a non-zero interaction at least at 1-$\sigma$ level. Note that the combined posterior distribution Planck+BAO+$H_0$+SNe+Ly$\alpha$ is not included here because it overlaps with the Planck+BAO+$H_0$ combination.  }
\end{figure*}

\begin{figure}
\includegraphics[width=0.45\textwidth]{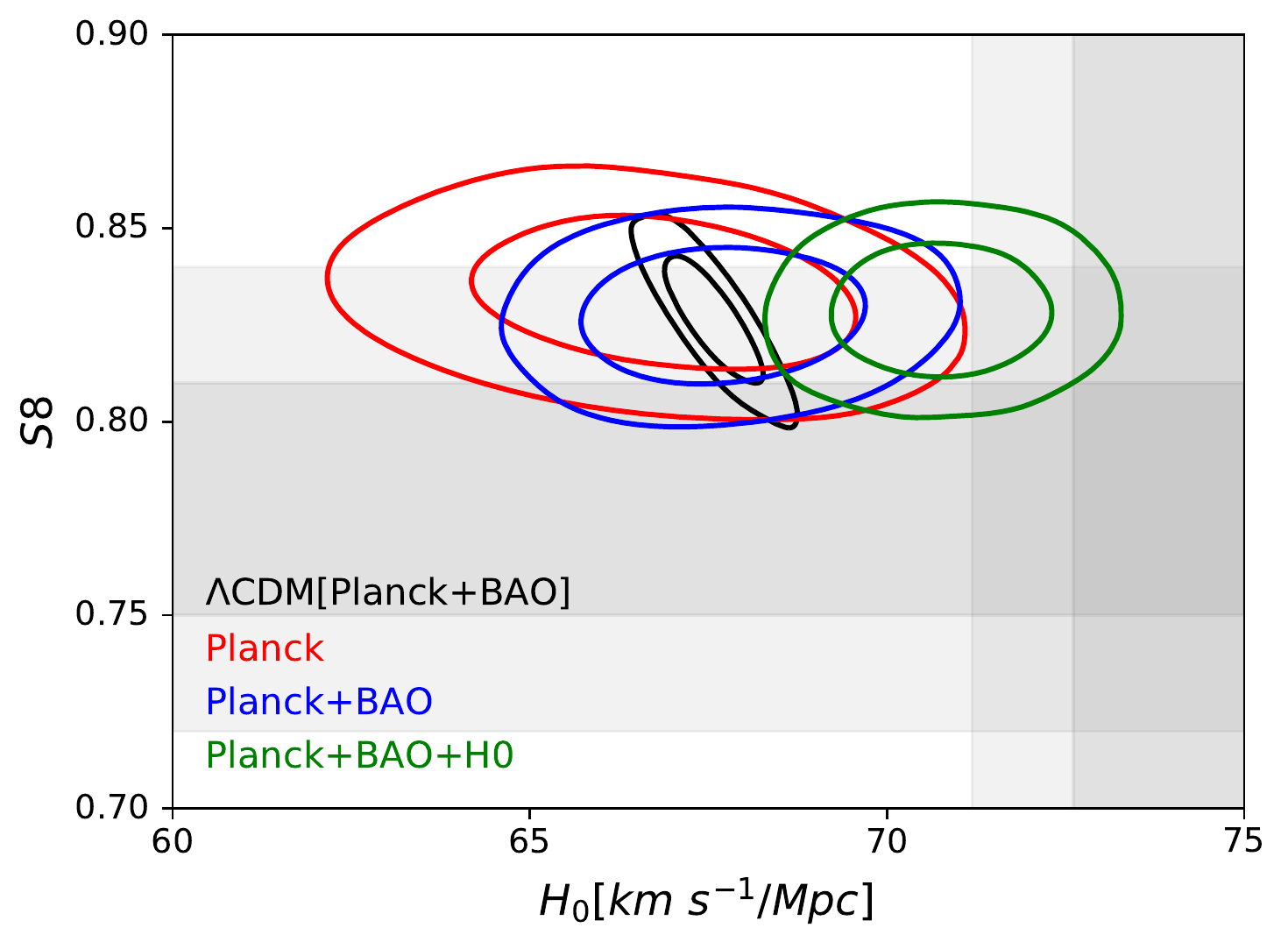}% Here is how to import EPS art
\caption{\label{fig:H0} 2D posteriors of $S_8$ and $H_0$. We compare posteriors in the self-interacting scenario using three data sets as follows Planck(red), Planck+BAO (blue), and Planck+BAO+H$_0$ (green). The posterior of $\Lambda$CDM is shown in black with the complete data set Planck+BAO+$H_0$+SNe+Ly-$\alpha$. The gray bands correspond to the 2$\sigma$ measurements for $S_8=\sigma_8(\Omega_m/0.3)^{0.5}=0.78\pm0.03$\cite{Hikage2019} and H$_0=74.03\pm 1.42$ km s$^{-1}$ Mpc$^{-1}$\cite{Riess2019ApJ}. The $\Lambda$CDM model (black) presents a tension of $4.3\sigma$ between the Planck+BAO and the local measurement. While the model with interactions reduces the tension to $3.2\sigma$ given the same data (blue).}
\end{figure}

In order to compare our model with the observations we run MCMC chains using the code MontePython \cite{Montepython2013,Brinckmann2019_montepython} with a modified version of CLASS \cite{class2, class4}. The model varies the usual $\Lambda$CDM parameters $\Omega_b$, $\Omega_{cdm}$, $\theta_s$, $A_s$, $n_s$ and $\tau_{reio}$ plus three extra parameters related with the neutrino dynamics which include the sum of the neutrino masses, $\sum m_\nu$, and the interaction strength $g_{\rm eff}$, as well as $N_{\rm ur}$ corresponding to a possible extra ultra relativistic component in the early Universe, with $N_{\rm eff}=3.046+N_{\rm ur}$.
The constraints obtained in the parameters are summarized in table \ref{tab:table3}, and in figures \ref{fig:triangle}-\ref{fig:H0}.
These figures were created with the GetDist package \cite{Lewis:2019xzd}.

In Fig. \ref{fig:triangle} we can see that the combination of CMB+BAO+$H_0$ data prefers a positive $g_{\rm eff}$ at least at one $\sigma$ level. This result mirrors the one obtained in Ref. \cite{Kreisch2020} for neutrino interactions with a massive mediator, where a maximum likelihood is found for a positive interaction strength. In that case, there is a second maximum close to zero strength, giving rise to two distinctive solutions preferred by the data, one with and one without interaction. In the case of the present work, where a light mediator drives the interaction, no second maximum is found close to $g_{\rm eff}=0$. Moreover, we can arguably expect that future data may prefer a positive interaction with a higher statistical significance.

We observe interesting correlations among the parameters.
For instance, when using any of the datasets, the correlation of the effective coupling, $g_{\rm eff}$,
with $H_0$ and with $N_{\rm eff}$ is positive and moderate, which means that a larger coupling value is associated with slightly larger $H_0$ and $N_{\rm eff}$ values.
While the correlation with the neutrino mass is rather weak.
Thus, this tells us that forcing the model to take a larger $H_0$-value can be partially compensated with a larger $g_{\rm eff}$-value.
Therefore, local measurements, which favor larger values for $H_0$ lead to larger values for the neutrino coupling.

Passing by, we also observe that when using Planck data only, the correlation between $\sum m_{\nu}$ and $H_0$ and also with $g_{\rm eff}$ is negative.
However, this negative correlation disappears when including BAO (and $H_0$) data.
Thus, having a larger neutrino mass does not prevent $H_0$ from taking larger values when the whole data pool is considered.
Notice that this effect is not unique to this model, and it mainly depends on the dataset selection \cite{Hou_2014}.

\begin{table*}
\caption{\label{tab:table3}Constraints on the cosmological parameters in the self-interacting neutrino scenario for different data combinations. We quote 68\% credible intervals, except for upper bounds, which are 95\%. For comparison purposes, the $\Delta \rm{AIC}$ values between the $\Lambda$CDM and the NSI models are included. The criterion favors the interacting model when $H_0$ data is included.}
\begin{ruledtabular}
\begin{tabular}{ccccc}
 %&\multicolumn{2}{c}{$D_{4h}^1$}&\multicolumn{2}{c}{$D_{4h}^5$}\\
 Parameter&Planck &Planck+BAO &Planck+ BAO+$H_0$
&Planck+BAO+$H_0$+SNe+Ly-$\alpha$\\ 
\hspace{1mm}
 $\Omega_bh^2\times10^2$ & $2.229\pm0.024$&$2.238\pm0.019$ &$2.271\pm0.017$&$2.272\pm0.016$\\ \hspace{1mm}
 $\Omega_{\rm cdm}h^2$ &$0.119\pm0.037$&$0.119\pm0.037$&$0.126\pm0.031$&$0.126\pm0.029$\\
 $100\theta_{s}$&$1.0422\pm0.0007$&$1.0422\pm0.0006$&$1.0413\pm0.0005$&$1.0413\pm0.0005$\\
 $\ln10^{10}A_{s }$&$3.042\pm0.019$&$3.044\pm0.017$&$3.061\pm0.016$&$3.061\pm0.016$\\
  $n_s$&$0.964\pm0.009$&$0.967\pm0.009$&$0.983\pm0.007$&$0.983\pm0.007$\\
 $\tau_{\rm reio}\times10^2$ &$5.54\pm0.83$&$5.72\pm0.78$&$5.97\pm0.77$&$6.01\pm0.80$\\
 $\sum m_{\nu}$ [eV] &$<0.23$&$<0.12$&$<0.12$&$<0.10$\\
 $g_{\rm eff}\times10^7$ &$<1.94$&$<1.97$&$<2.27$&$<2.30$\\
 $H_{0}$ &$66.9\pm1.9$&$67.7\pm1.4$&$70.8\pm1.1$&$70.9\pm1.0$\\
 $N_{\rm eff}$ &$3.00\pm0.24$&$3.03\pm0.22$&$3.49\pm0.19$&$3.50\pm0.17$\\
 $\sigma_8$ &$0.807\pm0.018$&$0.814\pm0.012$&$0.832\pm0.011$&$0.833\pm0.010$\\
 $\Delta \rm{AIC}$ & $4.49$  & $5.76$      &  $-3.88$  & $-5.20$           \\
\end{tabular}
\end{ruledtabular}
\end{table*}

Among the possible solutions, self-interacting neutrinos do not free-stream, and thus, their phase shift effect on the acoustic peaks can be suppressed, this can be partially compensated with a larger $N_{\rm eff}$-value and ultimately, due to their strong correlation, a larger value of $H_0$ is preferred by the data. This reduces, the tension in $H_0$ between the Planck and SH0ES estimates as can be seen in Figs. \ref{fig:triangle} and \ref{fig:H0}. The tension in $H_0$ between both Planck + BAO data and the SH0ES estimate is $4.3\sigma$ for the $\Lambda$CDM model, while it is only $3.2\sigma$ for the interactive neutrino model. In Fig. \ref{fig:H0} we show that the reduction in the $H_0$ tension doesn't introduce a tension on the parameter $S_8$.
Furthermore, as increasing neutrino masses would lower $H_0$, our bounds on $\Sigma m_\nu$ are consistent with those found by Planck collaboration~\cite{Aghanim2018Planck} and result tighter when $H_0$, SNe, and Ly-$\alpha$ data are included, as expected.

Adding new physics to the base $\Lambda$CDM model usually brings with it new degrees of freedom that can significantly improve the agreement of the model to data, and reduce tensions. However, this procedure can lead to overfitted models with nonphysical parameters. Some statistical tools have been created to evaluate whether the improvement in the fitting justifies the added parameters. Here we employ one such tool, known as the Akaike Information Criterion (AIC) defined as \cite{AIC, doi:10.1177/0049124104268644},
\begin{equation}
\rm{AIC}= -2\ln \mathcal{L}_{max}+2n_{free}\, ,
\end{equation}
where $ \mathcal{L}_{\rm max}$ is the maximum likelihood that can be obtained within the model, $\rm n_{free}$  the number of parameters. Between a family of models, the one with a smaller AIC value is preferred by the data. The addition of new parameters $\rm n_{free}$ is penalized but can be compensated if the more complex model has a much higher likelihood value. To compare the NSI model we use a base $\Lambda$CDM model with two massless and one massive neutrino with a fixed mass of $m_\nu=0.06$ eV. The lack of nonstandard interaction as well as the fixed mass and $N_{\rm eff}$ results in a model with 3 degrees of freedom less than the NSI model. We consider  $\Delta \rm{AIC}= \rm{AIC}_{\rm{NSI}}-\rm{AIC}_{\rm{\Lambda CDM}}$, a positive value for this quantity favors the base model while a negative value favors the NSI model.
From table \ref{tab:table3} we can see that $\Delta \rm{AIC}$ values are in the range $[4, 7]$ which means that the NSI model has statistical support although $\Lambda$CDM is preferred we can also note that the local $H_0$ data used together with Planck and BAO favors the interaction model. This result is encouraging as it justifies the inclusion of the interaction. For the larger set of data corresponding to Planck, BAO, $H_0$, Supernovae, and Ly-$\alpha$, the information criterion prefers the interaction model with an even larger margin.

%%%%%%%%%%%%%%%%%%%%%%%%%%%%%%%%%%%%%%%%%%%%%%%%%%%%%%%%%%%%%%%%%%%%%%%%%%%%%%%%%%%%%%
\section{\label{sec:conclusions} Summary and conclusions}
%%%%%%%%%%%%%%%%%%%%%%%%%%%%%%%%%%%%%%%%%%%%%%%%%%%%%%%%%%%%%%%%%%%%%%%%%%%%%%%%%%%%%%

In this work, we studied the cosmological effects of a nonstandard interaction between massive neutrinos mediated by a light particle. Our work generalizes similar studies like those in Refs. \cite{Forastieri2019} and \cite{Kreisch2020}, the former employing earlier Planck data and assuming a massless neutrino, and the latter using a heavy mediator.
For our purpose, we used a modified version of the Boltzmann solver CLASS \cite{class2,class4} to include the neutrino self-interactions and the Markov chain Monte Carlo code MontePython \cite{Montepython2013,Brinckmann2019_montepython} to obtain the parameter constraints.
Furthermore, we used several data sets including cosmic microwave background (CMB), baryonic acoustic oscillations (BAO), Ly-$\alpha$ forest, and local measurements of $H_0$.

The constrictions on the base--$\Lambda$CDM parameters reported in table \ref{tab:table3} are in good agreement with the known Planck 2018 results \cite{Planck:2018nkj}. The different data combinations are consistent with a null interaction at a 95\% confidence level, but with a peak at positive  $g_{\rm eff}$ which is more prominent when data from Planck + BAO combine with local data from $H_0$. 
In all these cases the magnitude of this interaction is about 
 $g_{\rm eff} \alt 2\times 10^{-7}$ (see table \ref{tab:table3}). 

We should stress that although our bound from only Planck data is slightly tighter than the one reported in Ref.~\cite{Forastieri2019} for interaction between massless neutrinos, it softens once additional cosmological data is included. 

Bounds on neutrino masses are consistent with previously known results, tightening a bit when additional data to Planck and BAO is considered. A larger $N_{\rm eff}$ is preferred by our analysis with all data sets which accompanies a larger $H_0$. The latter, however, does not suffice to resolve the $H_0$ tension. 

During this analysis, we have deliberately ignored the neutrino decay/annihilation into a lighter neutrino state and/or the very light scalar particle.
These processes may relax the neutrino mass bound \cite{Beacom2004,Hannestad2005,Chacko2020,Escudero2020relaxing,Esteban2021,Barenboim_2021,abellan2021improved}, while is not clear if this will have an important role in the determination of the $g_{\rm eff}$-constraint or on the $H_0$ tension, we plan to pursue this in a future project. 

The Akaike Information Criterion favors slightly the interactive model when data from Planck and $H_0$ combine. This results from the reduction in the tension between the local measurement of $H_0$ and its derivation from Planck + BAO data, from $4.3\sigma$ for the $\Lambda$CDM model to $3.2\sigma$ for the interactive neutrino model.

\acknowledgments
The authors thankfully acknowledge Cl\'uster de Superc\'omputo Xiuhcoatl (Cinvestav) for the allocation of computer resources.
Work partially supported by Conacyt, Mexico, under FORDECYT-PRONACES grant No. 490769.  G.G.A is grateful to FORDECYT PRONACES-CONACYT for partial support of the present research under Grant CF-MG-2558591.
We thank an anonymous referee for their useful comments that led to an improvement of our paper.

\bibliography{references}% Produces the bibliography via BibTeX.

\begin{thebibliography}{76}%
\makeatletter
\providecommand \@ifxundefined [1]{%
 \@ifx{#1\undefined}
}%
\providecommand \@ifnum [1]{%
 \ifnum #1\expandafter \@firstoftwo
 \else \expandafter \@secondoftwo
 \fi
}%
\providecommand \@ifx [1]{%
 \ifx #1\expandafter \@firstoftwo
 \else \expandafter \@secondoftwo
 \fi
}%
\providecommand \natexlab [1]{#1}%
\providecommand \enquote  [1]{``#1''}%
\providecommand \bibnamefont  [1]{#1}%
\providecommand \bibfnamefont [1]{#1}%
\providecommand \citenamefont [1]{#1}%
\providecommand \href@noop [0]{\@secondoftwo}%
\providecommand \href [0]{\begingroup \@sanitize@url \@href}%
\providecommand \@href[1]{\@@startlink{#1}\@@href}%
\providecommand \@@href[1]{\endgroup#1\@@endlink}%
\providecommand \@sanitize@url [0]{\catcode `\\12\catcode `\$12\catcode
  `\&12\catcode `\#12\catcode `\^12\catcode `\_12\catcode `\%12\relax}%
\providecommand \@@startlink[1]{}%
\providecommand \@@endlink[0]{}%
\providecommand \url  [0]{\begingroup\@sanitize@url \@url }%
\providecommand \@url [1]{\endgroup\@href {#1}{\urlprefix }}%
\providecommand \urlprefix  [0]{URL }%
\providecommand \Eprint [0]{\href }%
\providecommand \doibase [0]{http://dx.doi.org/}%
\providecommand \selectlanguage [0]{\@gobble}%
\providecommand \bibinfo  [0]{\@secondoftwo}%
\providecommand \bibfield  [0]{\@secondoftwo}%
\providecommand \translation [1]{[#1]}%
\providecommand \BibitemOpen [0]{}%
\providecommand \bibitemStop [0]{}%
\providecommand \bibitemNoStop [0]{.\EOS\space}%
\providecommand \EOS [0]{\spacefactor3000\relax}%
\providecommand \BibitemShut  [1]{\csname bibitem#1\endcsname}%
\let\auto@bib@innerbib\@empty
%</preamble>
\bibitem [{\citenamefont {Bashinsky}\ and\ \citenamefont
  {Seljak}(2004)}]{Bashinsky2004}%
  \BibitemOpen
  \bibfield  {author} {\bibinfo {author} {\bibfnamefont {S.}~\bibnamefont
  {Bashinsky}}\ and\ \bibinfo {author} {\bibfnamefont {U.~c.~v.}\ \bibnamefont
  {Seljak}},\ }\href {\doibase 10.1103/PhysRevD.69.083002} {\bibfield
  {journal} {\bibinfo  {journal} {\prd}\ }\textbf {\bibinfo {volume} {69}},\
  \bibinfo {pages} {083002} (\bibinfo {year} {2004})}\BibitemShut {NoStop}%
\bibitem [{\citenamefont {Follin}\ \emph {et~al.}(2015)\citenamefont {Follin},
  \citenamefont {Knox}, \citenamefont {Millea},\ and\ \citenamefont
  {Pan}}]{Follin2015}%
  \BibitemOpen
  \bibfield  {author} {\bibinfo {author} {\bibfnamefont {B.}~\bibnamefont
  {Follin}}, \bibinfo {author} {\bibfnamefont {L.}~\bibnamefont {Knox}},
  \bibinfo {author} {\bibfnamefont {M.}~\bibnamefont {Millea}}, \ and\ \bibinfo
  {author} {\bibfnamefont {Z.}~\bibnamefont {Pan}},\ }\href {\doibase
  10.1103/PhysRevLett.115.091301} {\bibfield  {journal} {\bibinfo  {journal}
  {\prl}\ }\textbf {\bibinfo {volume} {115}},\ \bibinfo {pages} {091301}
  (\bibinfo {year} {2015})}\BibitemShut {NoStop}%
\bibitem [{\citenamefont {Baumann}\ \emph {et~al.}(2019)\citenamefont
  {Baumann}, \citenamefont {Beutler}, \citenamefont {Flauger}, \citenamefont
  {Green}, \citenamefont {Slosar}, \citenamefont {Vargas-Maga{\~n}a},
  \citenamefont {Wallisch},\ and\ \citenamefont
  {Y{\`e}che}}]{baumann2019first}%
  \BibitemOpen
  \bibfield  {author} {\bibinfo {author} {\bibfnamefont {D.}~\bibnamefont
  {Baumann}}, \bibinfo {author} {\bibfnamefont {F.}~\bibnamefont {Beutler}},
  \bibinfo {author} {\bibfnamefont {R.}~\bibnamefont {Flauger}}, \bibinfo
  {author} {\bibfnamefont {D.}~\bibnamefont {Green}}, \bibinfo {author}
  {\bibfnamefont {A.}~\bibnamefont {Slosar}}, \bibinfo {author} {\bibfnamefont
  {M.}~\bibnamefont {Vargas-Maga{\~n}a}}, \bibinfo {author} {\bibfnamefont
  {B.}~\bibnamefont {Wallisch}}, \ and\ \bibinfo {author} {\bibfnamefont
  {C.}~\bibnamefont {Y{\`e}che}},\ }\href {\doibase 10.1038/s41567-019-0435-6}
  {\bibfield  {journal} {\bibinfo  {journal} {Nature Physics}\ }\textbf
  {\bibinfo {volume} {15}},\ \bibinfo {pages} {465} (\bibinfo {year}
  {2019})}\BibitemShut {NoStop}%
\bibitem [{\citenamefont {Mangano}\ \emph {et~al.}(2005)\citenamefont
  {Mangano}, \citenamefont {Miele}, \citenamefont {Pastor}, \citenamefont
  {Pinto}, \citenamefont {Pisanti},\ and\ \citenamefont
  {Serpico}}]{Mangano2005}%
  \BibitemOpen
  \bibfield  {author} {\bibinfo {author} {\bibfnamefont {G.}~\bibnamefont
  {Mangano}}, \bibinfo {author} {\bibfnamefont {G.}~\bibnamefont {Miele}},
  \bibinfo {author} {\bibfnamefont {S.}~\bibnamefont {Pastor}}, \bibinfo
  {author} {\bibfnamefont {T.}~\bibnamefont {Pinto}}, \bibinfo {author}
  {\bibfnamefont {O.}~\bibnamefont {Pisanti}}, \ and\ \bibinfo {author}
  {\bibfnamefont {P.~D.}\ \bibnamefont {Serpico}},\ }\href {\doibase
  https://doi.org/10.1016/j.nuclphysb.2005.09.041} {\bibfield  {journal}
  {\bibinfo  {journal} {Nuclear Physics B}\ }\textbf {\bibinfo {volume}
  {729}},\ \bibinfo {pages} {221 } (\bibinfo {year} {2005})}\BibitemShut
  {NoStop}%
\bibitem [{\citenamefont {de~Salas}\ and\ \citenamefont
  {Pastor}(2016)}]{deSalas2016}%
  \BibitemOpen
  \bibfield  {author} {\bibinfo {author} {\bibfnamefont {P.~F.}\ \bibnamefont
  {de~Salas}}\ and\ \bibinfo {author} {\bibfnamefont {S.}~\bibnamefont
  {Pastor}},\ }\href {\doibase 10.1088/1475-7516/2016/07/051} {\bibfield
  {journal} {\bibinfo  {journal} {Journal of Cosmology and Astroparticle
  Physics}\ }\textbf {\bibinfo {volume} {2016}},\ \bibinfo {pages} {051}
  (\bibinfo {year} {2016})}\BibitemShut {NoStop}%
\bibitem [{\citenamefont {Abenza}(2020)}]{Escudero2020}%
  \BibitemOpen
  \bibfield  {author} {\bibinfo {author} {\bibfnamefont {M.~E.}\ \bibnamefont
  {Abenza}},\ }\href {\doibase 10.1088/1475-7516/2020/05/048} {\bibfield
  {journal} {\bibinfo  {journal} {Journal of Cosmology and Astroparticle
  Physics}\ }\textbf {\bibinfo {volume} {2020}},\ \bibinfo {pages} {048}
  (\bibinfo {year} {2020})}\BibitemShut {NoStop}%
\bibitem [{\citenamefont {Akita}\ and\ \citenamefont
  {Yamaguchi}(2020)}]{Akita2020}%
  \BibitemOpen
  \bibfield  {author} {\bibinfo {author} {\bibfnamefont {K.}~\bibnamefont
  {Akita}}\ and\ \bibinfo {author} {\bibfnamefont {M.}~\bibnamefont
  {Yamaguchi}},\ }\href {\doibase 10.1088/1475-7516/2020/08/012} {\bibfield
  {journal} {\bibinfo  {journal} {Journal of Cosmology and Astroparticle
  Physics}\ }\textbf {\bibinfo {volume} {2020}},\ \bibinfo {pages} {012}
  (\bibinfo {year} {2020})}\BibitemShut {NoStop}%
\bibitem [{\citenamefont {Froustey}\ \emph {et~al.}(2020)\citenamefont
  {Froustey}, \citenamefont {Pitrou},\ and\ \citenamefont
  {Volpe}}]{froustey2020neff}%
  \BibitemOpen
  \bibfield  {author} {\bibinfo {author} {\bibfnamefont {J.}~\bibnamefont
  {Froustey}}, \bibinfo {author} {\bibfnamefont {C.}~\bibnamefont {Pitrou}}, \
  and\ \bibinfo {author} {\bibfnamefont {M.~C.}\ \bibnamefont {Volpe}},\ }\href
  {\doibase 10.1088/1475-7516/2020/12/015} {\bibfield  {journal} {\bibinfo
  {journal} {Journal of Cosmology and Astroparticle Physics}\ }\textbf
  {\bibinfo {volume} {2020}},\ \bibinfo {pages} {015} (\bibinfo {year}
  {2020})}\BibitemShut {NoStop}%
\bibitem [{\citenamefont {{Aghanim, N.}}\ \emph {et~al.}(2020)\citenamefont
  {{Aghanim, N.}} \emph {et~al.}}]{Aghanim2018Planck}%
  \BibitemOpen
  \bibfield  {author} {\bibinfo {author} {\bibnamefont {{Aghanim, N.}}} \emph
  {et~al.} (\bibinfo {collaboration} {Planck Collaboration}),\ }\href {\doibase
  10.1051/0004-6361/201833910} {\bibfield  {journal} {\bibinfo  {journal}
  {A\&A}\ }\textbf {\bibinfo {volume} {641}},\ \bibinfo {pages} {A6} (\bibinfo
  {year} {2020})}\BibitemShut {NoStop}%
\bibitem [{\citenamefont {Giusarma}\ \emph {et~al.}(2016)\citenamefont
  {Giusarma}, \citenamefont {Gerbino}, \citenamefont {Mena}, \citenamefont
  {Vagnozzi}, \citenamefont {Ho},\ and\ \citenamefont {Freese}}]{Guisarma2016}%
  \BibitemOpen
  \bibfield  {author} {\bibinfo {author} {\bibfnamefont {E.}~\bibnamefont
  {Giusarma}}, \bibinfo {author} {\bibfnamefont {M.}~\bibnamefont {Gerbino}},
  \bibinfo {author} {\bibfnamefont {O.}~\bibnamefont {Mena}}, \bibinfo {author}
  {\bibfnamefont {S.}~\bibnamefont {Vagnozzi}}, \bibinfo {author}
  {\bibfnamefont {S.}~\bibnamefont {Ho}}, \ and\ \bibinfo {author}
  {\bibfnamefont {K.}~\bibnamefont {Freese}},\ }\href {\doibase
  10.1103/PhysRevD.94.083522} {\bibfield  {journal} {\bibinfo  {journal} {Phys.
  Rev. D}\ }\textbf {\bibinfo {volume} {94}},\ \bibinfo {pages} {083522}
  (\bibinfo {year} {2016})}\BibitemShut {NoStop}%
\bibitem [{\citenamefont {Loureiro}\ \emph {et~al.}(2019)\citenamefont
  {Loureiro}, \citenamefont {Cuceu}, \citenamefont {Abdalla}, \citenamefont
  {Moraes}, \citenamefont {Whiteway}, \citenamefont {McLeod}, \citenamefont
  {Balan}, \citenamefont {Lahav}, \citenamefont {Benoit-L\'evy}, \citenamefont
  {Manera}, \citenamefont {Rollins},\ and\ \citenamefont
  {Xavier}}]{Loureiro2019}%
  \BibitemOpen
  \bibfield  {author} {\bibinfo {author} {\bibfnamefont {A.}~\bibnamefont
  {Loureiro}}, \bibinfo {author} {\bibfnamefont {A.}~\bibnamefont {Cuceu}},
  \bibinfo {author} {\bibfnamefont {F.~B.}\ \bibnamefont {Abdalla}}, \bibinfo
  {author} {\bibfnamefont {B.}~\bibnamefont {Moraes}}, \bibinfo {author}
  {\bibfnamefont {L.}~\bibnamefont {Whiteway}}, \bibinfo {author}
  {\bibfnamefont {M.}~\bibnamefont {McLeod}}, \bibinfo {author} {\bibfnamefont
  {S.~T.}\ \bibnamefont {Balan}}, \bibinfo {author} {\bibfnamefont
  {O.}~\bibnamefont {Lahav}}, \bibinfo {author} {\bibfnamefont
  {A.}~\bibnamefont {Benoit-L\'evy}}, \bibinfo {author} {\bibfnamefont
  {M.}~\bibnamefont {Manera}}, \bibinfo {author} {\bibfnamefont {R.~P.}\
  \bibnamefont {Rollins}}, \ and\ \bibinfo {author} {\bibfnamefont {H.~S.}\
  \bibnamefont {Xavier}},\ }\href {\doibase 10.1103/PhysRevLett.123.081301}
  {\bibfield  {journal} {\bibinfo  {journal} {Phys. Rev. Lett.}\ }\textbf
  {\bibinfo {volume} {123}},\ \bibinfo {pages} {081301} (\bibinfo {year}
  {2019})}\BibitemShut {NoStop}%
\bibitem [{\citenamefont {Vagnozzi}\ \emph {et~al.}(2017)\citenamefont
  {Vagnozzi}, \citenamefont {Giusarma}, \citenamefont {Mena}, \citenamefont
  {Freese}, \citenamefont {Gerbino}, \citenamefont {Ho},\ and\ \citenamefont
  {Lattanzi}}]{Vagnozzi2017}%
  \BibitemOpen
  \bibfield  {author} {\bibinfo {author} {\bibfnamefont {S.}~\bibnamefont
  {Vagnozzi}}, \bibinfo {author} {\bibfnamefont {E.}~\bibnamefont {Giusarma}},
  \bibinfo {author} {\bibfnamefont {O.}~\bibnamefont {Mena}}, \bibinfo {author}
  {\bibfnamefont {K.}~\bibnamefont {Freese}}, \bibinfo {author} {\bibfnamefont
  {M.}~\bibnamefont {Gerbino}}, \bibinfo {author} {\bibfnamefont
  {S.}~\bibnamefont {Ho}}, \ and\ \bibinfo {author} {\bibfnamefont
  {M.}~\bibnamefont {Lattanzi}},\ }\href {\doibase 10.1103/PhysRevD.96.123503}
  {\bibfield  {journal} {\bibinfo  {journal} {Phys. Rev. D}\ }\textbf {\bibinfo
  {volume} {96}},\ \bibinfo {pages} {123503} (\bibinfo {year}
  {2017})}\BibitemShut {NoStop}%
\bibitem [{\citenamefont {Brinckmann}\ \emph {et~al.}(2019)\citenamefont
  {Brinckmann}, \citenamefont {Hooper}, \citenamefont {Archidiacono},
  \citenamefont {Lesgourgues},\ and\ \citenamefont
  {Sprenger}}]{Brinckmann2019}%
  \BibitemOpen
  \bibfield  {author} {\bibinfo {author} {\bibfnamefont {T.}~\bibnamefont
  {Brinckmann}}, \bibinfo {author} {\bibfnamefont {D.~C.}\ \bibnamefont
  {Hooper}}, \bibinfo {author} {\bibfnamefont {M.}~\bibnamefont
  {Archidiacono}}, \bibinfo {author} {\bibfnamefont {J.}~\bibnamefont
  {Lesgourgues}}, \ and\ \bibinfo {author} {\bibfnamefont {T.}~\bibnamefont
  {Sprenger}},\ }\href {\doibase 10.1088/1475-7516/2019/01/059} {\bibfield
  {journal} {\bibinfo  {journal} {Journal of Cosmology and Astroparticle
  Physics}\ }\textbf {\bibinfo {volume} {2019}},\ \bibinfo {pages} {059}
  (\bibinfo {year} {2019})}\BibitemShut {NoStop}%
\bibitem [{\citenamefont {Di~Valentino}\ \emph {et~al.}(2021)\citenamefont
  {Di~Valentino}, \citenamefont {Gariazzo},\ and\ \citenamefont
  {Mena}}]{DiValentino2021}%
  \BibitemOpen
  \bibfield  {author} {\bibinfo {author} {\bibfnamefont {E.}~\bibnamefont
  {Di~Valentino}}, \bibinfo {author} {\bibfnamefont {S.}~\bibnamefont
  {Gariazzo}}, \ and\ \bibinfo {author} {\bibfnamefont {O.}~\bibnamefont
  {Mena}},\ }\href {\doibase 10.1103/PhysRevD.104.083504} {\bibfield  {journal}
  {\bibinfo  {journal} {Phys. Rev. D}\ }\textbf {\bibinfo {volume} {104}},\
  \bibinfo {pages} {083504} (\bibinfo {year} {2021})}\BibitemShut {NoStop}%
\bibitem [{\citenamefont {Bell}\ \emph {et~al.}(2006)\citenamefont {Bell},
  \citenamefont {Pierpaoli},\ and\ \citenamefont {Sigurdson}}]{Bell2006}%
  \BibitemOpen
  \bibfield  {author} {\bibinfo {author} {\bibfnamefont {N.~F.}\ \bibnamefont
  {Bell}}, \bibinfo {author} {\bibfnamefont {E.}~\bibnamefont {Pierpaoli}}, \
  and\ \bibinfo {author} {\bibfnamefont {K.}~\bibnamefont {Sigurdson}},\ }\href
  {\doibase 10.1103/PhysRevD.73.063523} {\bibfield  {journal} {\bibinfo
  {journal} {Phys. Rev. D}\ }\textbf {\bibinfo {volume} {73}},\ \bibinfo
  {pages} {063523} (\bibinfo {year} {2006})}\BibitemShut {NoStop}%
\bibitem [{\citenamefont {Archidiacono}\ and\ \citenamefont
  {Hannestad}(2014)}]{Archidiacono2014}%
  \BibitemOpen
  \bibfield  {author} {\bibinfo {author} {\bibfnamefont {M.}~\bibnamefont
  {Archidiacono}}\ and\ \bibinfo {author} {\bibfnamefont {S.}~\bibnamefont
  {Hannestad}},\ }\href {\doibase 10.1088/1475-7516/2014/07/046} {\bibfield
  {journal} {\bibinfo  {journal} {Journal of Cosmology and Astroparticle
  Physics}\ }\textbf {\bibinfo {volume} {2014}},\ \bibinfo {pages} {046}
  (\bibinfo {year} {2014})}\BibitemShut {NoStop}%
\bibitem [{\citenamefont {Cyr-Racine}\ and\ \citenamefont
  {Sigurdson}(2014)}]{CyrRacine2014}%
  \BibitemOpen
  \bibfield  {author} {\bibinfo {author} {\bibfnamefont {F.-Y.}\ \bibnamefont
  {Cyr-Racine}}\ and\ \bibinfo {author} {\bibfnamefont {K.}~\bibnamefont
  {Sigurdson}},\ }\href {\doibase 10.1103/PhysRevD.90.123533} {\bibfield
  {journal} {\bibinfo  {journal} {Phys. Rev. D}\ }\textbf {\bibinfo {volume}
  {90}},\ \bibinfo {pages} {123533} (\bibinfo {year} {2014})}\BibitemShut
  {NoStop}%
\bibitem [{\citenamefont {Lancaster}\ \emph {et~al.}(2017)\citenamefont
  {Lancaster}, \citenamefont {Cyr-Racine}, \citenamefont {Knox},\ and\
  \citenamefont {Pan}}]{Lancaster2017}%
  \BibitemOpen
  \bibfield  {author} {\bibinfo {author} {\bibfnamefont {L.}~\bibnamefont
  {Lancaster}}, \bibinfo {author} {\bibfnamefont {F.-Y.}\ \bibnamefont
  {Cyr-Racine}}, \bibinfo {author} {\bibfnamefont {L.}~\bibnamefont {Knox}}, \
  and\ \bibinfo {author} {\bibfnamefont {Z.}~\bibnamefont {Pan}},\ }\href
  {\doibase 10.1088/1475-7516/2017/07/033} {\bibfield  {journal} {\bibinfo
  {journal} {Journal of Cosmology and Astroparticle Physics}\ }\textbf
  {\bibinfo {volume} {2017}},\ \bibinfo {pages} {033} (\bibinfo {year}
  {2017})}\BibitemShut {NoStop}%
\bibitem [{\citenamefont {Kreisch}\ \emph {et~al.}(2020)\citenamefont
  {Kreisch}, \citenamefont {Cyr-Racine},\ and\ \citenamefont
  {Dor\'e}}]{Kreisch2020}%
  \BibitemOpen
  \bibfield  {author} {\bibinfo {author} {\bibfnamefont {C.~D.}\ \bibnamefont
  {Kreisch}}, \bibinfo {author} {\bibfnamefont {F.-Y.}\ \bibnamefont
  {Cyr-Racine}}, \ and\ \bibinfo {author} {\bibfnamefont {O.}~\bibnamefont
  {Dor\'e}},\ }\href {\doibase 10.1103/PhysRevD.101.123505} {\bibfield
  {journal} {\bibinfo  {journal} {Phys. Rev. D}\ }\textbf {\bibinfo {volume}
  {101}},\ \bibinfo {pages} {123505} (\bibinfo {year} {2020})}\BibitemShut
  {NoStop}%
\bibitem [{\citenamefont {Park}\ \emph {et~al.}(2019)\citenamefont {Park},
  \citenamefont {Kreisch}, \citenamefont {Dunkley}, \citenamefont
  {Hadzhiyska},\ and\ \citenamefont {Cyr-Racine}}]{Park2019}%
  \BibitemOpen
  \bibfield  {author} {\bibinfo {author} {\bibfnamefont {M.}~\bibnamefont
  {Park}}, \bibinfo {author} {\bibfnamefont {C.~D.}\ \bibnamefont {Kreisch}},
  \bibinfo {author} {\bibfnamefont {J.}~\bibnamefont {Dunkley}}, \bibinfo
  {author} {\bibfnamefont {B.}~\bibnamefont {Hadzhiyska}}, \ and\ \bibinfo
  {author} {\bibfnamefont {F.-Y.}\ \bibnamefont {Cyr-Racine}},\ }\href
  {\doibase 10.1103/PhysRevD.100.063524} {\bibfield  {journal} {\bibinfo
  {journal} {Phys. Rev. D}\ }\textbf {\bibinfo {volume} {100}},\ \bibinfo
  {pages} {063524} (\bibinfo {year} {2019})}\BibitemShut {NoStop}%
\bibitem [{\citenamefont {Choudhury}\ \emph {et~al.}(2021)\citenamefont
  {Choudhury}, \citenamefont {Hannestad},\ and\ \citenamefont
  {Tram}}]{Choudhury2021}%
  \BibitemOpen
  \bibfield  {author} {\bibinfo {author} {\bibfnamefont {S.~R.}\ \bibnamefont
  {Choudhury}}, \bibinfo {author} {\bibfnamefont {S.}~\bibnamefont
  {Hannestad}}, \ and\ \bibinfo {author} {\bibfnamefont {T.}~\bibnamefont
  {Tram}},\ }\href {\doibase 10.1088/1475-7516/2021/03/084} {\bibfield
  {journal} {\bibinfo  {journal} {Journal of Cosmology and Astroparticle
  Physics}\ }\textbf {\bibinfo {volume} {2021}},\ \bibinfo {pages} {084}
  (\bibinfo {year} {2021})}\BibitemShut {NoStop}%
\bibitem [{\citenamefont {Verde}\ \emph {et~al.}(2019)\citenamefont {Verde},
  \citenamefont {Treu},\ and\ \citenamefont {Riess}}]{Verde2019}%
  \BibitemOpen
  \bibfield  {author} {\bibinfo {author} {\bibfnamefont {L.}~\bibnamefont
  {Verde}}, \bibinfo {author} {\bibfnamefont {T.}~\bibnamefont {Treu}}, \ and\
  \bibinfo {author} {\bibfnamefont {A.~G.}\ \bibnamefont {Riess}},\ }\href
  {\doibase 10.1038/s41550-019-0902-0} {\bibfield  {journal} {\bibinfo
  {journal} {Nature Astronomy}\ }\textbf {\bibinfo {volume} {3}},\ \bibinfo
  {pages} {891} (\bibinfo {year} {2019})}\BibitemShut {NoStop}%
\bibitem [{\citenamefont {Aiola}\ \emph {et~al.}(2020)\citenamefont {Aiola}
  \emph {et~al.}}]{ACT020}%
  \BibitemOpen
  \bibfield  {author} {\bibinfo {author} {\bibfnamefont {S.}~\bibnamefont
  {Aiola}} \emph {et~al.},\ }\href {\doibase 10.1088/1475-7516/2020/12/047}
  {\bibfield  {journal} {\bibinfo  {journal} {Journal of Cosmology and
  Astroparticle Physics}\ }\textbf {\bibinfo {volume} {2020}},\ \bibinfo
  {pages} {047} (\bibinfo {year} {2020})}\BibitemShut {NoStop}%
\bibitem [{\citenamefont {Abbott}\ and\ \citenamefont { and the
  South Pole Telescope Collaborations)}(2018)}]{DES2018}%
  \BibitemOpen
  \bibfield  {author} {\bibinfo {author} {\bibfnamefont {T.~M.~C.}\
  \bibnamefont {Abbott}}\ and\ \bibinfo {author} {\bibfnamefont
  {A.}~\bibnamefont { and the South Pole Telescope Collaborations)}},\
  }\href {\doibase 10.1093/mnras/sty1939} {\bibfield  {journal} {\bibinfo
  {journal} {Monthly Notices of the Royal Astronomical Society}\ }\textbf
  {\bibinfo {volume} {480}},\ \bibinfo {pages} {3879} (\bibinfo {year}
  {2018})},\ \Eprint
  {http://arxiv.org/abs/https://academic.oup.com/mnras/article-pdf/480/3/3879/25520162/sty1939.pdf}
  {https://academic.oup.com/mnras/article-pdf/480/3/3879/25520162/sty1939.pdf}
  \BibitemShut {NoStop}%
\bibitem [{\citenamefont {Riess}\ \emph {et~al.}(2019)\citenamefont {Riess},
  \citenamefont {Casertano}, \citenamefont {Yuan}, \citenamefont {Macri},\ and\
  \citenamefont {Scolnic}}]{Riess2019ApJ}%
  \BibitemOpen
  \bibfield  {author} {\bibinfo {author} {\bibfnamefont {A.~G.}\ \bibnamefont
  {Riess}}, \bibinfo {author} {\bibfnamefont {S.}~\bibnamefont {Casertano}},
  \bibinfo {author} {\bibfnamefont {W.}~\bibnamefont {Yuan}}, \bibinfo {author}
  {\bibfnamefont {L.~M.}\ \bibnamefont {Macri}}, \ and\ \bibinfo {author}
  {\bibfnamefont {D.}~\bibnamefont {Scolnic}},\ }\href {\doibase
  10.3847/1538-4357/ab1422} {\bibfield  {journal} {\bibinfo  {journal} {The
  Astrophysical Journal}\ }\textbf {\bibinfo {volume} {876}},\ \bibinfo {pages}
  {85} (\bibinfo {year} {2019})}\BibitemShut {NoStop}%
\bibitem [{\citenamefont {Wong}\ \emph {et~al.}(2020)\citenamefont {Wong} \emph
  {et~al.}}]{H0licow2020}%
  \BibitemOpen
  \bibfield  {author} {\bibinfo {author} {\bibfnamefont {K.~C.}\ \bibnamefont
  {Wong}} \emph {et~al.} (\bibinfo {collaboration} {H0liCOW Collaboration}),\
  }\href {\doibase 10.1093/mnras/stz3094} {\bibfield  {journal} {\bibinfo
  {journal} {Monthly Notices of the Royal Astronomical Society}\ } (\bibinfo
  {year} {2020}),\ 10.1093/mnras/stz3094}\BibitemShut {NoStop}%
\bibitem [{\citenamefont {Heurtier}\ and\ \citenamefont
  {Zhang}(2017)}]{Heurtier2017}%
  \BibitemOpen
  \bibfield  {author} {\bibinfo {author} {\bibfnamefont {L.}~\bibnamefont
  {Heurtier}}\ and\ \bibinfo {author} {\bibfnamefont {Y.}~\bibnamefont
  {Zhang}},\ }\href {\doibase 10.1088/1475-7516/2017/02/042} {\bibfield
  {journal} {\bibinfo  {journal} {Journal of Cosmology and Astroparticle
  Physics}\ }\textbf {\bibinfo {volume} {2017}},\ \bibinfo {pages} {042}
  (\bibinfo {year} {2017})}\BibitemShut {NoStop}%
\bibitem [{\citenamefont {Huang}\ \emph {et~al.}(2018)\citenamefont {Huang},
  \citenamefont {Ohlsson},\ and\ \citenamefont {Zhou}}]{Huang2018}%
  \BibitemOpen
  \bibfield  {author} {\bibinfo {author} {\bibfnamefont {G.-y.}\ \bibnamefont
  {Huang}}, \bibinfo {author} {\bibfnamefont {T.}~\bibnamefont {Ohlsson}}, \
  and\ \bibinfo {author} {\bibfnamefont {S.}~\bibnamefont {Zhou}},\ }\href
  {\doibase 10.1103/PhysRevD.97.075009} {\bibfield  {journal} {\bibinfo
  {journal} {Phys. Rev. D}\ }\textbf {\bibinfo {volume} {97}},\ \bibinfo
  {pages} {075009} (\bibinfo {year} {2018})}\BibitemShut {NoStop}%
\bibitem [{\citenamefont {Blinov}\ \emph {et~al.}(2019)\citenamefont {Blinov},
  \citenamefont {Kelly}, \citenamefont {Krnjaic},\ and\ \citenamefont
  {McDermott}}]{Blinov2019}%
  \BibitemOpen
  \bibfield  {author} {\bibinfo {author} {\bibfnamefont {N.}~\bibnamefont
  {Blinov}}, \bibinfo {author} {\bibfnamefont {K.~J.}\ \bibnamefont {Kelly}},
  \bibinfo {author} {\bibfnamefont {G.}~\bibnamefont {Krnjaic}}, \ and\
  \bibinfo {author} {\bibfnamefont {S.~D.}\ \bibnamefont {McDermott}},\ }\href
  {\doibase 10.1103/PhysRevLett.123.191102} {\bibfield  {journal} {\bibinfo
  {journal} {Phys. Rev. Lett.}\ }\textbf {\bibinfo {volume} {123}},\ \bibinfo
  {pages} {191102} (\bibinfo {year} {2019})}\BibitemShut {NoStop}%
\bibitem [{\citenamefont {Brune}\ and\ \citenamefont
  {P\"as}(2019)}]{Brune2019}%
  \BibitemOpen
  \bibfield  {author} {\bibinfo {author} {\bibfnamefont {T.}~\bibnamefont
  {Brune}}\ and\ \bibinfo {author} {\bibfnamefont {H.}~\bibnamefont {P\"as}},\
  }\href {\doibase 10.1103/PhysRevD.99.096005} {\bibfield  {journal} {\bibinfo
  {journal} {Physical Review D}\ }\textbf {\bibinfo {volume} {99}},\ \bibinfo
  {pages} {96005} (\bibinfo {year} {2019})}\BibitemShut {NoStop}%
\bibitem [{\citenamefont {Brinckmann}\ \emph {et~al.}(2021)\citenamefont
  {Brinckmann}, \citenamefont {Chang},\ and\ \citenamefont
  {LoVerde}}]{brinckmann2020self}%
  \BibitemOpen
  \bibfield  {author} {\bibinfo {author} {\bibfnamefont {T.}~\bibnamefont
  {Brinckmann}}, \bibinfo {author} {\bibfnamefont {J.~H.}\ \bibnamefont
  {Chang}}, \ and\ \bibinfo {author} {\bibfnamefont {M.}~\bibnamefont
  {LoVerde}},\ }\href {\doibase 10.1103/PhysRevD.104.063523} {\bibfield
  {journal} {\bibinfo  {journal} {Phys. Rev. D}\ }\textbf {\bibinfo {volume}
  {104}},\ \bibinfo {pages} {063523} (\bibinfo {year} {2021})}\BibitemShut
  {NoStop}%
\bibitem [{\citenamefont {Mazumdar}\ \emph {et~al.}(2020)\citenamefont
  {Mazumdar}, \citenamefont {Mohanty},\ and\ \citenamefont
  {Parashari}}]{mazumdar2020flavour}%
  \BibitemOpen
  \bibfield  {author} {\bibinfo {author} {\bibfnamefont {A.}~\bibnamefont
  {Mazumdar}}, \bibinfo {author} {\bibfnamefont {S.}~\bibnamefont {Mohanty}}, \
  and\ \bibinfo {author} {\bibfnamefont {P.}~\bibnamefont {Parashari}},\
  }\href@noop {} {\bibfield  {journal} {\bibinfo  {journal} {arXiv preprint
  arXiv:2011.13685}\ } (\bibinfo {year} {2020})}\BibitemShut {NoStop}%
\bibitem [{\citenamefont {Das}\ and\ \citenamefont {Ghosh}(2021)}]{Das2021}%
  \BibitemOpen
  \bibfield  {author} {\bibinfo {author} {\bibfnamefont {A.}~\bibnamefont
  {Das}}\ and\ \bibinfo {author} {\bibfnamefont {S.}~\bibnamefont {Ghosh}},\
  }\href {\doibase 10.1088/1475-7516/2021/07/038} {\bibfield  {journal}
  {\bibinfo  {journal} {Journal of Cosmology and Astroparticle Physics}\
  }\textbf {\bibinfo {volume} {2021}},\ \bibinfo {pages} {038} (\bibinfo {year}
  {2021})}\BibitemShut {NoStop}%
\bibitem [{\citenamefont {Schöneberg}\ \emph {et~al.}(2019)\citenamefont
  {Schöneberg}, \citenamefont {Lesgourgues},\ and\ \citenamefont
  {Hooper}}]{Schoneberg2019}%
  \BibitemOpen
  \bibfield  {author} {\bibinfo {author} {\bibfnamefont {N.}~\bibnamefont
  {Schöneberg}}, \bibinfo {author} {\bibfnamefont {J.}~\bibnamefont
  {Lesgourgues}}, \ and\ \bibinfo {author} {\bibfnamefont {D.~C.}\ \bibnamefont
  {Hooper}},\ }\href {\doibase 10.1088/1475-7516/2019/10/029} {\bibfield
  {journal} {\bibinfo  {journal} {Journal of Cosmology and Astroparticle
  Physics}\ }\textbf {\bibinfo {volume} {2019}},\ \bibinfo {pages} {029}
  (\bibinfo {year} {2019})}\BibitemShut {NoStop}%
\bibitem [{\citenamefont {Huang}\ and\ \citenamefont
  {Rodejohann}(2021)}]{Huang2021}%
  \BibitemOpen
  \bibfield  {author} {\bibinfo {author} {\bibfnamefont {G.-y.}\ \bibnamefont
  {Huang}}\ and\ \bibinfo {author} {\bibfnamefont {W.}~\bibnamefont
  {Rodejohann}},\ }\href {\doibase 10.1103/PhysRevD.103.123007} {\bibfield
  {journal} {\bibinfo  {journal} {Phys. Rev. D}\ }\textbf {\bibinfo {volume}
  {103}},\ \bibinfo {pages} {123007} (\bibinfo {year} {2021})}\BibitemShut
  {NoStop}%
\bibitem [{\citenamefont {Brdar}\ \emph {et~al.}(2020)\citenamefont {Brdar},
  \citenamefont {Lindner}, \citenamefont {Vogl},\ and\ \citenamefont
  {Xu}}]{Brdar2020}%
  \BibitemOpen
  \bibfield  {author} {\bibinfo {author} {\bibfnamefont {V.}~\bibnamefont
  {Brdar}}, \bibinfo {author} {\bibfnamefont {M.}~\bibnamefont {Lindner}},
  \bibinfo {author} {\bibfnamefont {S.}~\bibnamefont {Vogl}}, \ and\ \bibinfo
  {author} {\bibfnamefont {X.-J.}\ \bibnamefont {Xu}},\ }\href {\doibase
  10.1103/PhysRevD.101.115001} {\bibfield  {journal} {\bibinfo  {journal}
  {Phys. Rev. D}\ }\textbf {\bibinfo {volume} {101}},\ \bibinfo {pages}
  {115001} (\bibinfo {year} {2020})}\BibitemShut {NoStop}%
\bibitem [{\citenamefont {Esteban}\ \emph {et~al.}(2021)\citenamefont
  {Esteban}, \citenamefont {Pandey}, \citenamefont {Brdar},\ and\ \citenamefont
  {Beacom}}]{Esteban2021_probing}%
  \BibitemOpen
  \bibfield  {author} {\bibinfo {author} {\bibfnamefont {I.}~\bibnamefont
  {Esteban}}, \bibinfo {author} {\bibfnamefont {S.}~\bibnamefont {Pandey}},
  \bibinfo {author} {\bibfnamefont {V.}~\bibnamefont {Brdar}}, \ and\ \bibinfo
  {author} {\bibfnamefont {J.~F.}\ \bibnamefont {Beacom}},\ }\href {\doibase
  10.1103/PhysRevD.104.123014} {\bibfield  {journal} {\bibinfo  {journal}
  {Phys. Rev. D}\ }\textbf {\bibinfo {volume} {104}},\ \bibinfo {pages}
  {123014} (\bibinfo {year} {2021})}\BibitemShut {NoStop}%
\bibitem [{\citenamefont {Aguilar}\ \emph {et~al.}(2001)\citenamefont
  {Aguilar}, \citenamefont {Auerbach}, \citenamefont {Burman}, \citenamefont
  {Caldwell}, \citenamefont {Church}, \citenamefont {Cochran}, \citenamefont
  {Donahue}, \citenamefont {Fazely}, \citenamefont {Garvey}, \citenamefont
  {Gunasingha}, \citenamefont {Imlay}, \citenamefont {Louis}, \citenamefont
  {Majkic}, \citenamefont {Malik}, \citenamefont {Metcalf}, \citenamefont
  {Mills}, \citenamefont {Sandberg}, \citenamefont {Smith}, \citenamefont
  {Stancu}, \citenamefont {Sung}, \citenamefont {Tayloe}, \citenamefont
  {VanDalen}, \citenamefont {Vernon}, \citenamefont {Wadia}, \citenamefont
  {White},\ and\ \citenamefont {Yellin}}]{LSND2001}%
  \BibitemOpen
  \bibfield  {author} {\bibinfo {author} {\bibfnamefont {A.}~\bibnamefont
  {Aguilar}}, \bibinfo {author} {\bibfnamefont {L.~B.}\ \bibnamefont
  {Auerbach}}, \bibinfo {author} {\bibfnamefont {R.~L.}\ \bibnamefont
  {Burman}}, \bibinfo {author} {\bibfnamefont {D.~O.}\ \bibnamefont
  {Caldwell}}, \bibinfo {author} {\bibfnamefont {E.~D.}\ \bibnamefont
  {Church}}, \bibinfo {author} {\bibfnamefont {A.~K.}\ \bibnamefont {Cochran}},
  \bibinfo {author} {\bibfnamefont {J.~B.}\ \bibnamefont {Donahue}}, \bibinfo
  {author} {\bibfnamefont {A.}~\bibnamefont {Fazely}}, \bibinfo {author}
  {\bibfnamefont {G.~T.}\ \bibnamefont {Garvey}}, \bibinfo {author}
  {\bibfnamefont {R.~M.}\ \bibnamefont {Gunasingha}}, \bibinfo {author}
  {\bibfnamefont {R.}~\bibnamefont {Imlay}}, \bibinfo {author} {\bibfnamefont
  {W.~C.}\ \bibnamefont {Louis}}, \bibinfo {author} {\bibfnamefont
  {R.}~\bibnamefont {Majkic}}, \bibinfo {author} {\bibfnamefont
  {A.}~\bibnamefont {Malik}}, \bibinfo {author} {\bibfnamefont
  {W.}~\bibnamefont {Metcalf}}, \bibinfo {author} {\bibfnamefont {G.~B.}\
  \bibnamefont {Mills}}, \bibinfo {author} {\bibfnamefont {V.}~\bibnamefont
  {Sandberg}}, \bibinfo {author} {\bibfnamefont {D.}~\bibnamefont {Smith}},
  \bibinfo {author} {\bibfnamefont {I.}~\bibnamefont {Stancu}}, \bibinfo
  {author} {\bibfnamefont {M.}~\bibnamefont {Sung}}, \bibinfo {author}
  {\bibfnamefont {R.}~\bibnamefont {Tayloe}}, \bibinfo {author} {\bibfnamefont
  {G.~J.}\ \bibnamefont {VanDalen}}, \bibinfo {author} {\bibfnamefont
  {W.}~\bibnamefont {Vernon}}, \bibinfo {author} {\bibfnamefont
  {N.}~\bibnamefont {Wadia}}, \bibinfo {author} {\bibfnamefont {D.~H.}\
  \bibnamefont {White}}, \ and\ \bibinfo {author} {\bibfnamefont
  {S.}~\bibnamefont {Yellin}} (\bibinfo {collaboration} {LSND Collaboration}),\
  }\href {\doibase 10.1103/PhysRevD.64.112007} {\bibfield  {journal} {\bibinfo
  {journal} {Phys. Rev. D}\ }\textbf {\bibinfo {volume} {64}},\ \bibinfo
  {pages} {112007} (\bibinfo {year} {2001})}\BibitemShut {NoStop}%
\bibitem [{\citenamefont {Aguilar-Arevalo}\ \emph {et~al.}(2018)\citenamefont
  {Aguilar-Arevalo}, \citenamefont {Brown}, \citenamefont {Bugel},
  \citenamefont {Cheng}, \citenamefont {Conrad}, \citenamefont {Cooper} \emph
  {et~al.}}]{MiniBoone2018}%
  \BibitemOpen
  \bibfield  {author} {\bibinfo {author} {\bibfnamefont {A.~A.}\ \bibnamefont
  {Aguilar-Arevalo}}, \bibinfo {author} {\bibfnamefont {B.~C.}\ \bibnamefont
  {Brown}}, \bibinfo {author} {\bibfnamefont {L.}~\bibnamefont {Bugel}},
  \bibinfo {author} {\bibfnamefont {G.}~\bibnamefont {Cheng}}, \bibinfo
  {author} {\bibfnamefont {J.~M.}\ \bibnamefont {Conrad}}, \bibinfo {author}
  {\bibfnamefont {R.~L.}\ \bibnamefont {Cooper}},  \emph {et~al.} (\bibinfo
  {collaboration} {MiniBooNE Collaboration}),\ }\href {\doibase
  10.1103/PhysRevLett.121.221801} {\bibfield  {journal} {\bibinfo  {journal}
  {Phys. Rev. Lett.}\ }\textbf {\bibinfo {volume} {121}},\ \bibinfo {pages}
  {221801} (\bibinfo {year} {2018})}\BibitemShut {NoStop}%
\bibitem [{\citenamefont {Palomares-Ruiz}\ \emph {et~al.}(2005)\citenamefont
  {Palomares-Ruiz}, \citenamefont {Pascoli},\ and\ \citenamefont
  {Schwetz}}]{Palomares2005}%
  \BibitemOpen
  \bibfield  {author} {\bibinfo {author} {\bibfnamefont {S.}~\bibnamefont
  {Palomares-Ruiz}}, \bibinfo {author} {\bibfnamefont {S.}~\bibnamefont
  {Pascoli}}, \ and\ \bibinfo {author} {\bibfnamefont {T.}~\bibnamefont
  {Schwetz}},\ }\href {\doibase 10.1088/1126-6708/2005/09/048} {\bibfield
  {journal} {\bibinfo  {journal} {Journal of High Energy Physics}\ }\textbf
  {\bibinfo {volume} {2005}},\ \bibinfo {pages} {048} (\bibinfo {year}
  {2005})}\BibitemShut {NoStop}%
\bibitem [{\citenamefont {de~Gouv{\^e}a}\ \emph {et~al.}(2020)\citenamefont
  {de~Gouv{\^e}a}, \citenamefont {Peres}, \citenamefont {Prakash},\ and\
  \citenamefont {Stenico}}]{deGouvea2020}%
  \BibitemOpen
  \bibfield  {author} {\bibinfo {author} {\bibfnamefont {A.}~\bibnamefont
  {de~Gouv{\^e}a}}, \bibinfo {author} {\bibfnamefont {O.~L.~G.}\ \bibnamefont
  {Peres}}, \bibinfo {author} {\bibfnamefont {S.}~\bibnamefont {Prakash}}, \
  and\ \bibinfo {author} {\bibfnamefont {G.~V.}\ \bibnamefont {Stenico}},\
  }\href {\doibase 10.1007/JHEP07(2020)141} {\bibfield  {journal} {\bibinfo
  {journal} {Journal of High Energy Physics}\ }\textbf {\bibinfo {volume}
  {2020}},\ \bibinfo {pages} {141} (\bibinfo {year} {2020})}\BibitemShut
  {NoStop}%
\bibitem [{\citenamefont {Abratenko}\ \emph {et~al.}(2021)\citenamefont
  {Abratenko}, \citenamefont {An}, \citenamefont {Anthony}, \citenamefont
  {Arellano}, \citenamefont {Asaadi}, \citenamefont {Ashkenazi}, \citenamefont
  {Balasubramanian}, \citenamefont {Baller}, \citenamefont {Barnes},
  \citenamefont {Barr} \emph {et~al.}}]{abratenko2021search}%
  \BibitemOpen
  \bibfield  {author} {\bibinfo {author} {\bibfnamefont {P.}~\bibnamefont
  {Abratenko}}, \bibinfo {author} {\bibfnamefont {R.}~\bibnamefont {An}},
  \bibinfo {author} {\bibfnamefont {J.}~\bibnamefont {Anthony}}, \bibinfo
  {author} {\bibfnamefont {L.}~\bibnamefont {Arellano}}, \bibinfo {author}
  {\bibfnamefont {J.}~\bibnamefont {Asaadi}}, \bibinfo {author} {\bibfnamefont
  {A.}~\bibnamefont {Ashkenazi}}, \bibinfo {author} {\bibfnamefont
  {S.}~\bibnamefont {Balasubramanian}}, \bibinfo {author} {\bibfnamefont
  {B.}~\bibnamefont {Baller}}, \bibinfo {author} {\bibfnamefont
  {C.}~\bibnamefont {Barnes}}, \bibinfo {author} {\bibfnamefont
  {G.}~\bibnamefont {Barr}},  \emph {et~al.},\ }\href@noop {} {\bibfield
  {journal} {\bibinfo  {journal} {arXiv preprint arXiv:2110.14054}\ } (\bibinfo
  {year} {2021})}\BibitemShut {NoStop}%
\bibitem [{\citenamefont {Beacom}\ \emph {et~al.}(2004)\citenamefont {Beacom},
  \citenamefont {Bell},\ and\ \citenamefont {Dodelson}}]{Beacom2004}%
  \BibitemOpen
  \bibfield  {author} {\bibinfo {author} {\bibfnamefont {J.~F.}\ \bibnamefont
  {Beacom}}, \bibinfo {author} {\bibfnamefont {N.~F.}\ \bibnamefont {Bell}}, \
  and\ \bibinfo {author} {\bibfnamefont {S.}~\bibnamefont {Dodelson}},\ }\href
  {\doibase 10.1103/PhysRevLett.93.121302} {\bibfield  {journal} {\bibinfo
  {journal} {Phys. Rev. Lett.}\ }\textbf {\bibinfo {volume} {93}},\ \bibinfo
  {pages} {121302} (\bibinfo {year} {2004})}\BibitemShut {NoStop}%
\bibitem [{\citenamefont {Hannestad}(2005)}]{Hannestad2005}%
  \BibitemOpen
  \bibfield  {author} {\bibinfo {author} {\bibfnamefont {S.}~\bibnamefont
  {Hannestad}},\ }\href {\doibase 10.1088/1475-7516/2005/02/011} {\bibfield
  {journal} {\bibinfo  {journal} {Journal of Cosmology and Astroparticle
  Physics}\ }\textbf {\bibinfo {volume} {2005}},\ \bibinfo {pages} {011}
  (\bibinfo {year} {2005})}\BibitemShut {NoStop}%
\bibitem [{\citenamefont {Chacko}\ \emph {et~al.}(2020)\citenamefont {Chacko},
  \citenamefont {Dev}, \citenamefont {Du}, \citenamefont {Poulin},\ and\
  \citenamefont {Tsai}}]{Chacko2020}%
  \BibitemOpen
  \bibfield  {author} {\bibinfo {author} {\bibfnamefont {Z.}~\bibnamefont
  {Chacko}}, \bibinfo {author} {\bibfnamefont {A.}~\bibnamefont {Dev}},
  \bibinfo {author} {\bibfnamefont {P.}~\bibnamefont {Du}}, \bibinfo {author}
  {\bibfnamefont {V.}~\bibnamefont {Poulin}}, \ and\ \bibinfo {author}
  {\bibfnamefont {Y.}~\bibnamefont {Tsai}},\ }\href {\doibase
  10.1007/JHEP04(2020)020} {\bibfield  {journal} {\bibinfo  {journal} {Journal
  of High Energy Physics}\ }\textbf {\bibinfo {volume} {2020}},\ \bibinfo
  {pages} {20} (\bibinfo {year} {2020})}\BibitemShut {NoStop}%
\bibitem [{\citenamefont {Escudero}\ \emph {et~al.}(2020)\citenamefont
  {Escudero}, \citenamefont {Lopez-Pavon}, \citenamefont {Rius},\ and\
  \citenamefont {Sandner}}]{Escudero2020relaxing}%
  \BibitemOpen
  \bibfield  {author} {\bibinfo {author} {\bibfnamefont {M.}~\bibnamefont
  {Escudero}}, \bibinfo {author} {\bibfnamefont {J.}~\bibnamefont
  {Lopez-Pavon}}, \bibinfo {author} {\bibfnamefont {N.}~\bibnamefont {Rius}}, \
  and\ \bibinfo {author} {\bibfnamefont {S.}~\bibnamefont {Sandner}},\ }\href
  {\doibase 10.1007/JHEP12(2020)119} {\bibfield  {journal} {\bibinfo  {journal}
  {Journal of High Energy Physics}\ }\textbf {\bibinfo {volume} {2020}},\
  \bibinfo {pages} {119} (\bibinfo {year} {2020})}\BibitemShut {NoStop}%
\bibitem [{\citenamefont {Esteban}\ and\ \citenamefont
  {Salvado}(2021)}]{Esteban2021}%
  \BibitemOpen
  \bibfield  {author} {\bibinfo {author} {\bibfnamefont {I.}~\bibnamefont
  {Esteban}}\ and\ \bibinfo {author} {\bibfnamefont {J.}~\bibnamefont
  {Salvado}},\ }\href {\doibase 10.1088/1475-7516/2021/05/036} {\bibfield
  {journal} {\bibinfo  {journal} {Journal of Cosmology and Astroparticle
  Physics}\ }\textbf {\bibinfo {volume} {2021}},\ \bibinfo {pages} {036}
  (\bibinfo {year} {2021})}\BibitemShut {NoStop}%
\bibitem [{\citenamefont {Barenboim}\ \emph {et~al.}(2021)\citenamefont
  {Barenboim}, \citenamefont {Chen}, \citenamefont {Hannestad}, \citenamefont
  {Oldengott}, \citenamefont {Tram},\ and\ \citenamefont
  {Wong}}]{Barenboim_2021}%
  \BibitemOpen
  \bibfield  {author} {\bibinfo {author} {\bibfnamefont {G.}~\bibnamefont
  {Barenboim}}, \bibinfo {author} {\bibfnamefont {J.~Z.}\ \bibnamefont {Chen}},
  \bibinfo {author} {\bibfnamefont {S.}~\bibnamefont {Hannestad}}, \bibinfo
  {author} {\bibfnamefont {I.~M.}\ \bibnamefont {Oldengott}}, \bibinfo {author}
  {\bibfnamefont {T.}~\bibnamefont {Tram}}, \ and\ \bibinfo {author}
  {\bibfnamefont {Y.~Y.}\ \bibnamefont {Wong}},\ }\href {\doibase
  10.1088/1475-7516/2021/03/087} {\bibfield  {journal} {\bibinfo  {journal}
  {Journal of Cosmology and Astroparticle Physics}\ }\textbf {\bibinfo {volume}
  {2021}},\ \bibinfo {pages} {087} (\bibinfo {year} {2021})}\BibitemShut
  {NoStop}%
\bibitem [{\citenamefont {Abell{\'a}n}\ \emph {et~al.}(2021)\citenamefont
  {Abell{\'a}n}, \citenamefont {Chacko}, \citenamefont {Dev}, \citenamefont
  {Du}, \citenamefont {Poulin},\ and\ \citenamefont
  {Tsai}}]{abellan2021improved}%
  \BibitemOpen
  \bibfield  {author} {\bibinfo {author} {\bibfnamefont {G.~F.}\ \bibnamefont
  {Abell{\'a}n}}, \bibinfo {author} {\bibfnamefont {Z.}~\bibnamefont {Chacko}},
  \bibinfo {author} {\bibfnamefont {A.}~\bibnamefont {Dev}}, \bibinfo {author}
  {\bibfnamefont {P.}~\bibnamefont {Du}}, \bibinfo {author} {\bibfnamefont
  {V.}~\bibnamefont {Poulin}}, \ and\ \bibinfo {author} {\bibfnamefont
  {Y.}~\bibnamefont {Tsai}},\ }\href@noop {} {\bibfield  {journal} {\bibinfo
  {journal} {arXiv preprint arXiv:2112.13862}\ } (\bibinfo {year}
  {2021})}\BibitemShut {NoStop}%
\bibitem [{\citenamefont {Venzor}\ \emph {et~al.}(2021)\citenamefont {Venzor},
  \citenamefont {P\'erez-Lorenzana},\ and\ \citenamefont
  {De-Santiago}}]{Venzor2021}%
  \BibitemOpen
  \bibfield  {author} {\bibinfo {author} {\bibfnamefont {J.}~\bibnamefont
  {Venzor}}, \bibinfo {author} {\bibfnamefont {A.}~\bibnamefont
  {P\'erez-Lorenzana}}, \ and\ \bibinfo {author} {\bibfnamefont
  {J.}~\bibnamefont {De-Santiago}},\ }\href {\doibase
  10.1103/PhysRevD.103.043534} {\bibfield  {journal} {\bibinfo  {journal}
  {Phys. Rev. D}\ }\textbf {\bibinfo {volume} {103}},\ \bibinfo {pages}
  {043534} (\bibinfo {year} {2021})}\BibitemShut {NoStop}%
\bibitem [{\citenamefont {Forastieri}\ \emph {et~al.}(2015)\citenamefont
  {Forastieri}, \citenamefont {Lattanzi},\ and\ \citenamefont
  {Natoli}}]{Forastieri2015}%
  \BibitemOpen
  \bibfield  {author} {\bibinfo {author} {\bibfnamefont {F.}~\bibnamefont
  {Forastieri}}, \bibinfo {author} {\bibfnamefont {M.}~\bibnamefont
  {Lattanzi}}, \ and\ \bibinfo {author} {\bibfnamefont {P.}~\bibnamefont
  {Natoli}},\ }\href {\doibase 10.1088/1475-7516/2015/07/014} {\bibfield
  {journal} {\bibinfo  {journal} {Journal of Cosmology and Astroparticle
  Physics}\ }\textbf {\bibinfo {volume} {2015}},\ \bibinfo {pages} {014}
  (\bibinfo {year} {2015})}\BibitemShut {NoStop}%
\bibitem [{\citenamefont {Forastieri}\ \emph {et~al.}(2019)\citenamefont
  {Forastieri}, \citenamefont {Lattanzi},\ and\ \citenamefont
  {Natoli}}]{Forastieri2019}%
  \BibitemOpen
  \bibfield  {author} {\bibinfo {author} {\bibfnamefont {F.}~\bibnamefont
  {Forastieri}}, \bibinfo {author} {\bibfnamefont {M.}~\bibnamefont
  {Lattanzi}}, \ and\ \bibinfo {author} {\bibfnamefont {P.}~\bibnamefont
  {Natoli}},\ }\href {\doibase 10.1103/PhysRevD.100.103526} {\bibfield
  {journal} {\bibinfo  {journal} {Phys. Rev. D}\ }\textbf {\bibinfo {volume}
  {100}},\ \bibinfo {pages} {103526} (\bibinfo {year} {2019})}\BibitemShut
  {NoStop}%
\bibitem [{\citenamefont {Choi}\ \emph {et~al.}(2018)\citenamefont {Choi},
  \citenamefont {Chiang},\ and\ \citenamefont {LoVerde}}]{Choi_2018}%
  \BibitemOpen
  \bibfield  {author} {\bibinfo {author} {\bibfnamefont {G.}~\bibnamefont
  {Choi}}, \bibinfo {author} {\bibfnamefont {C.-T.}\ \bibnamefont {Chiang}}, \
  and\ \bibinfo {author} {\bibfnamefont {M.}~\bibnamefont {LoVerde}},\ }\href
  {\doibase 10.1088/1475-7516/2018/06/044} {\bibfield  {journal} {\bibinfo
  {journal} {Journal of Cosmology and Astroparticle Physics}\ }\textbf
  {\bibinfo {volume} {2018}},\ \bibinfo {pages} {044} (\bibinfo {year}
  {2018})}\BibitemShut {NoStop}%
\bibitem [{\citenamefont {Baumann}\ \emph {et~al.}(2017)\citenamefont
  {Baumann}, \citenamefont {Green},\ and\ \citenamefont
  {Zaldarriaga}}]{Baumann_2017}%
  \BibitemOpen
  \bibfield  {author} {\bibinfo {author} {\bibfnamefont {D.}~\bibnamefont
  {Baumann}}, \bibinfo {author} {\bibfnamefont {D.}~\bibnamefont {Green}}, \
  and\ \bibinfo {author} {\bibfnamefont {M.}~\bibnamefont {Zaldarriaga}},\
  }\href {\doibase 10.1088/1475-7516/2017/11/007} {\bibfield  {journal}
  {\bibinfo  {journal} {Journal of Cosmology and Astroparticle Physics}\
  }\textbf {\bibinfo {volume} {2017}},\ \bibinfo {pages} {007} (\bibinfo {year}
  {2017})}\BibitemShut {NoStop}%
\bibitem [{\citenamefont {Oldengott}\ \emph {et~al.}(2015)\citenamefont
  {Oldengott}, \citenamefont {Rampf},\ and\ \citenamefont
  {Wong}}]{Oldengott2015}%
  \BibitemOpen
  \bibfield  {author} {\bibinfo {author} {\bibfnamefont {I.~M.}\ \bibnamefont
  {Oldengott}}, \bibinfo {author} {\bibfnamefont {C.}~\bibnamefont {Rampf}}, \
  and\ \bibinfo {author} {\bibfnamefont {Y.~Y.}\ \bibnamefont {Wong}},\ }\href
  {\doibase 10.1088/1475-7516/2015/04/016} {\bibfield  {journal} {\bibinfo
  {journal} {Journal of Cosmology and Astroparticle Physics}\ }\textbf
  {\bibinfo {volume} {2015}},\ \bibinfo {pages} {016} (\bibinfo {year}
  {2015})}\BibitemShut {NoStop}%
\bibitem [{\citenamefont {Oldengott}\ \emph {et~al.}(2017)\citenamefont
  {Oldengott}, \citenamefont {Tram}, \citenamefont {Rampf},\ and\ \citenamefont
  {Wong}}]{Oldengott2017}%
  \BibitemOpen
  \bibfield  {author} {\bibinfo {author} {\bibfnamefont {I.~M.}\ \bibnamefont
  {Oldengott}}, \bibinfo {author} {\bibfnamefont {T.}~\bibnamefont {Tram}},
  \bibinfo {author} {\bibfnamefont {C.}~\bibnamefont {Rampf}}, \ and\ \bibinfo
  {author} {\bibfnamefont {Y.~Y.}\ \bibnamefont {Wong}},\ }\href {\doibase
  10.1088/1475-7516/2017/11/027} {\bibfield  {journal} {\bibinfo  {journal}
  {Journal of Cosmology and Astroparticle Physics}\ }\textbf {\bibinfo {volume}
  {2017}},\ \bibinfo {pages} {027} (\bibinfo {year} {2017})}\BibitemShut
  {NoStop}%
\bibitem [{\citenamefont {Basb\o{}ll}\ \emph {et~al.}(2009)\citenamefont
  {Basb\o{}ll}, \citenamefont {Bjaelde}, \citenamefont {Hannestad},\ and\
  \citenamefont {Raffelt}}]{Basboll2009}%
  \BibitemOpen
  \bibfield  {author} {\bibinfo {author} {\bibfnamefont {A.}~\bibnamefont
  {Basb\o{}ll}}, \bibinfo {author} {\bibfnamefont {O.~E.}\ \bibnamefont
  {Bjaelde}}, \bibinfo {author} {\bibfnamefont {S.}~\bibnamefont {Hannestad}},
  \ and\ \bibinfo {author} {\bibfnamefont {G.~G.}\ \bibnamefont {Raffelt}},\
  }\href {\doibase 10.1103/PhysRevD.79.043512} {\bibfield  {journal} {\bibinfo
  {journal} {Phys. Rev. D}\ }\textbf {\bibinfo {volume} {79}},\ \bibinfo
  {pages} {043512} (\bibinfo {year} {2009})}\BibitemShut {NoStop}%
\bibitem [{\citenamefont {Lesgourgues}\ and\ \citenamefont
  {Pastor}(2006)}]{Lesgourgues:2006nd}%
  \BibitemOpen
  \bibfield  {author} {\bibinfo {author} {\bibfnamefont {J.}~\bibnamefont
  {Lesgourgues}}\ and\ \bibinfo {author} {\bibfnamefont {S.}~\bibnamefont
  {Pastor}},\ }\href {\doibase 10.1016/j.physrep.2006.04.001} {\bibfield
  {journal} {\bibinfo  {journal} {Phys. Rept.}\ }\textbf {\bibinfo {volume}
  {429}},\ \bibinfo {pages} {307} (\bibinfo {year} {2006})},\ \Eprint
  {http://arxiv.org/abs/astro-ph/0603494} {arXiv:astro-ph/0603494} \BibitemShut
  {NoStop}%
\bibitem [{\citenamefont {Ma}\ and\ \citenamefont
  {Bertschinger}(1995)}]{Ma:1995ey}%
  \BibitemOpen
  \bibfield  {author} {\bibinfo {author} {\bibfnamefont {C.-P.}\ \bibnamefont
  {Ma}}\ and\ \bibinfo {author} {\bibfnamefont {E.}~\bibnamefont
  {Bertschinger}},\ }\href {\doibase 10.1086/176550} {\bibfield  {journal}
  {\bibinfo  {journal} {Astrophys. J.}\ }\textbf {\bibinfo {volume} {455}},\
  \bibinfo {pages} {7} (\bibinfo {year} {1995})},\ \Eprint
  {http://arxiv.org/abs/astro-ph/9506072} {arXiv:astro-ph/9506072} \BibitemShut
  {NoStop}%
\bibitem [{\citenamefont {Hannestad}\ and\ \citenamefont
  {Scherrer}(2000)}]{Hannestad:2000gt}%
  \BibitemOpen
  \bibfield  {author} {\bibinfo {author} {\bibfnamefont {S.}~\bibnamefont
  {Hannestad}}\ and\ \bibinfo {author} {\bibfnamefont {R.~J.}\ \bibnamefont
  {Scherrer}},\ }\href {\doibase 10.1103/PhysRevD.62.043522} {\bibfield
  {journal} {\bibinfo  {journal} {Phys. Rev. D}\ }\textbf {\bibinfo {volume}
  {62}},\ \bibinfo {pages} {043522} (\bibinfo {year} {2000})},\ \Eprint
  {http://arxiv.org/abs/astro-ph/0003046} {arXiv:astro-ph/0003046} \BibitemShut
  {NoStop}%
\bibitem [{\citenamefont {Aghanim}\ \emph {et~al.}(2020)\citenamefont {Aghanim}
  \emph {et~al.}}]{Planck:2018nkj}%
  \BibitemOpen
  \bibfield  {author} {\bibinfo {author} {\bibfnamefont {N.}~\bibnamefont
  {Aghanim}} \emph {et~al.} (\bibinfo {collaboration} {Planck}),\ }\href
  {\doibase 10.1051/0004-6361/201833880} {\bibfield  {journal} {\bibinfo
  {journal} {Astron. Astrophys.}\ }\textbf {\bibinfo {volume} {641}},\ \bibinfo
  {pages} {A1} (\bibinfo {year} {2020})},\ \Eprint
  {http://arxiv.org/abs/1807.06205} {arXiv:1807.06205 [astro-ph.CO]}
  \BibitemShut {NoStop}%
\bibitem [{\citenamefont {Alam}\ \emph {et~al.}(2017)\citenamefont {Alam} \emph
  {et~al.}}]{BOSS:2016wmc}%
  \BibitemOpen
  \bibfield  {author} {\bibinfo {author} {\bibfnamefont {S.}~\bibnamefont
  {Alam}} \emph {et~al.} (\bibinfo {collaboration} {BOSS}),\ }\href {\doibase
  10.1093/mnras/stx721} {\bibfield  {journal} {\bibinfo  {journal} {Mon. Not.
  Roy. Astron. Soc.}\ }\textbf {\bibinfo {volume} {470}},\ \bibinfo {pages}
  {2617} (\bibinfo {year} {2017})},\ \Eprint {http://arxiv.org/abs/1607.03155}
  {arXiv:1607.03155 [astro-ph.CO]} \BibitemShut {NoStop}%
\bibitem [{\citenamefont {Bautista}\ \emph {et~al.}(2018)\citenamefont
  {Bautista} \emph {et~al.}}]{Bautista:2017wwp}%
  \BibitemOpen
  \bibfield  {author} {\bibinfo {author} {\bibfnamefont {J.~E.}\ \bibnamefont
  {Bautista}} \emph {et~al.},\ }\href {\doibase 10.3847/1538-4357/aacea5}
  {\bibfield  {journal} {\bibinfo  {journal} {Astrophys. J.}\ }\textbf
  {\bibinfo {volume} {863}},\ \bibinfo {pages} {110} (\bibinfo {year}
  {2018})},\ \Eprint {http://arxiv.org/abs/1712.08064} {arXiv:1712.08064
  [astro-ph.CO]} \BibitemShut {NoStop}%
\bibitem [{\citenamefont {du~Mas~des Bourboux}\ \emph
  {et~al.}(2017)\citenamefont {du~Mas~des Bourboux} \emph
  {et~al.}}]{duMasdesBourboux:2017mrl}%
  \BibitemOpen
  \bibfield  {author} {\bibinfo {author} {\bibfnamefont {H.}~\bibnamefont
  {du~Mas~des Bourboux}} \emph {et~al.},\ }\href {\doibase
  10.1051/0004-6361/201731731} {\bibfield  {journal} {\bibinfo  {journal}
  {Astron. Astrophys.}\ }\textbf {\bibinfo {volume} {608}},\ \bibinfo {pages}
  {A130} (\bibinfo {year} {2017})},\ \Eprint {http://arxiv.org/abs/1708.02225}
  {arXiv:1708.02225 [astro-ph.CO]} \BibitemShut {NoStop}%
\bibitem [{\citenamefont {Beutler}\ \emph {et~al.}(2011)\citenamefont
  {Beutler}, \citenamefont {Blake}, \citenamefont {Colless}, \citenamefont
  {Jones}, \citenamefont {Staveley-Smith}, \citenamefont {Campbell},
  \citenamefont {Parker}, \citenamefont {Saunders},\ and\ \citenamefont
  {Watson}}]{Beutler:2011hx}%
  \BibitemOpen
  \bibfield  {author} {\bibinfo {author} {\bibfnamefont {F.}~\bibnamefont
  {Beutler}}, \bibinfo {author} {\bibfnamefont {C.}~\bibnamefont {Blake}},
  \bibinfo {author} {\bibfnamefont {M.}~\bibnamefont {Colless}}, \bibinfo
  {author} {\bibfnamefont {D.~H.}\ \bibnamefont {Jones}}, \bibinfo {author}
  {\bibfnamefont {L.}~\bibnamefont {Staveley-Smith}}, \bibinfo {author}
  {\bibfnamefont {L.}~\bibnamefont {Campbell}}, \bibinfo {author}
  {\bibfnamefont {Q.}~\bibnamefont {Parker}}, \bibinfo {author} {\bibfnamefont
  {W.}~\bibnamefont {Saunders}}, \ and\ \bibinfo {author} {\bibfnamefont
  {F.}~\bibnamefont {Watson}},\ }\href {\doibase
  10.1111/j.1365-2966.2011.19250.x} {\bibfield  {journal} {\bibinfo  {journal}
  {Mon. Not. Roy. Astron. Soc.}\ }\textbf {\bibinfo {volume} {416}},\ \bibinfo
  {pages} {3017} (\bibinfo {year} {2011})},\ \Eprint
  {http://arxiv.org/abs/1106.3366} {arXiv:1106.3366 [astro-ph.CO]} \BibitemShut
  {NoStop}%
\bibitem [{\citenamefont {Ross}\ \emph {et~al.}(2015)\citenamefont {Ross},
  \citenamefont {Samushia}, \citenamefont {Howlett}, \citenamefont {Percival},
  \citenamefont {Burden},\ and\ \citenamefont {Manera}}]{Ross:2014qpa}%
  \BibitemOpen
  \bibfield  {author} {\bibinfo {author} {\bibfnamefont {A.~J.}\ \bibnamefont
  {Ross}}, \bibinfo {author} {\bibfnamefont {L.}~\bibnamefont {Samushia}},
  \bibinfo {author} {\bibfnamefont {C.}~\bibnamefont {Howlett}}, \bibinfo
  {author} {\bibfnamefont {W.~J.}\ \bibnamefont {Percival}}, \bibinfo {author}
  {\bibfnamefont {A.}~\bibnamefont {Burden}}, \ and\ \bibinfo {author}
  {\bibfnamefont {M.}~\bibnamefont {Manera}},\ }\href {\doibase
  10.1093/mnras/stv154} {\bibfield  {journal} {\bibinfo  {journal} {Mon. Not.
  Roy. Astron. Soc.}\ }\textbf {\bibinfo {volume} {449}},\ \bibinfo {pages}
  {835} (\bibinfo {year} {2015})},\ \Eprint {http://arxiv.org/abs/1409.3242}
  {arXiv:1409.3242 [astro-ph.CO]} \BibitemShut {NoStop}%
\bibitem [{\citenamefont {Scolnic}\ \emph {et~al.}(2018)\citenamefont {Scolnic}
  \emph {et~al.}}]{Pan-STARRS1:2017jku}%
  \BibitemOpen
  \bibfield  {author} {\bibinfo {author} {\bibfnamefont {D.~M.}\ \bibnamefont
  {Scolnic}} \emph {et~al.} (\bibinfo {collaboration} {Pan-STARRS1}),\ }\href
  {\doibase 10.3847/1538-4357/aab9bb} {\bibfield  {journal} {\bibinfo
  {journal} {Astrophys. J.}\ }\textbf {\bibinfo {volume} {859}},\ \bibinfo
  {pages} {101} (\bibinfo {year} {2018})},\ \Eprint
  {http://arxiv.org/abs/1710.00845} {arXiv:1710.00845 [astro-ph.CO]}
  \BibitemShut {NoStop}%
\bibitem [{\citenamefont {Audren}\ \emph {et~al.}(2013)\citenamefont {Audren},
  \citenamefont {Lesgourgues}, \citenamefont {Benabed},\ and\ \citenamefont
  {Prunet}}]{Montepython2013}%
  \BibitemOpen
  \bibfield  {author} {\bibinfo {author} {\bibfnamefont {B.}~\bibnamefont
  {Audren}}, \bibinfo {author} {\bibfnamefont {J.}~\bibnamefont {Lesgourgues}},
  \bibinfo {author} {\bibfnamefont {K.}~\bibnamefont {Benabed}}, \ and\
  \bibinfo {author} {\bibfnamefont {S.}~\bibnamefont {Prunet}},\ }\href
  {http://stacks.iop.org/1475-7516/2013/i=02/a=001} {\bibfield  {journal}
  {\bibinfo  {journal} {jcap}\ }\textbf {\bibinfo {volume} {2013}},\ \bibinfo
  {pages} {001} (\bibinfo {year} {2013})}\BibitemShut {NoStop}%
\bibitem [{\citenamefont {Brinckmann}\ and\ \citenamefont
  {Lesgourgues}(2019)}]{Brinckmann2019_montepython}%
  \BibitemOpen
  \bibfield  {author} {\bibinfo {author} {\bibfnamefont {T.}~\bibnamefont
  {Brinckmann}}\ and\ \bibinfo {author} {\bibfnamefont {J.}~\bibnamefont
  {Lesgourgues}},\ }\href {\doibase https://doi.org/10.1016/j.dark.2018.100260}
  {\bibfield  {journal} {\bibinfo  {journal} {Physics of the Dark Universe}\
  }\textbf {\bibinfo {volume} {24}},\ \bibinfo {pages} {100260} (\bibinfo
  {year} {2019})}\BibitemShut {NoStop}%
\bibitem [{\citenamefont {Blas}\ \emph {et~al.}(2011)\citenamefont {Blas},
  \citenamefont {Lesgourgues},\ and\ \citenamefont {Tram}}]{class2}%
  \BibitemOpen
  \bibfield  {author} {\bibinfo {author} {\bibfnamefont {D.}~\bibnamefont
  {Blas}}, \bibinfo {author} {\bibfnamefont {J.}~\bibnamefont {Lesgourgues}}, \
  and\ \bibinfo {author} {\bibfnamefont {T.}~\bibnamefont {Tram}},\ }\href
  {\doibase http://dx.doi.org/10.1088/1475-7516/2011/07/034} {\bibfield
  {journal} {\bibinfo  {journal} {Journal of Cosmology and Astroparticle
  Physics}\ }\textbf {\bibinfo {volume} {2011}},\ \bibinfo {pages} {034}
  (\bibinfo {year} {2011})}\BibitemShut {NoStop}%
\bibitem [{\citenamefont {Lesgourgues}\ and\ \citenamefont
  {Tram}(2011)}]{class4}%
  \BibitemOpen
  \bibfield  {author} {\bibinfo {author} {\bibfnamefont {J.}~\bibnamefont
  {Lesgourgues}}\ and\ \bibinfo {author} {\bibfnamefont {T.}~\bibnamefont
  {Tram}},\ }\href {\doibase 10.1088/1475-7516/2011/07/034} {\bibfield
  {journal} {\bibinfo  {journal} {Journal of Cosmology and Astroparticle
  Physics}\ }\textbf {\bibinfo {volume} {2011}},\ \bibinfo {pages} {032}
  (\bibinfo {year} {2011})}\BibitemShut {NoStop}%
\bibitem [{\citenamefont {Lewis}(2019)}]{Lewis:2019xzd}%
  \BibitemOpen
  \bibfield  {author} {\bibinfo {author} {\bibfnamefont {A.}~\bibnamefont
  {Lewis}},\ }\href@noop {} {\enquote {\bibinfo {title} {{GetDist: a Python
  package for analysing Monte Carlo samples}},}\ } (\bibinfo {year} {2019}),\
  \bibinfo {note} {https://getdist.readthedocs.io},\ \Eprint
  {http://arxiv.org/abs/1910.13970} {arXiv:1910.13970 [astro-ph.IM]}
  \BibitemShut {NoStop}%
%%CITATION = ARXIV:1910.13970;%%
\bibitem [{\citenamefont {Hou}\ \emph {et~al.}(2014)\citenamefont {Hou} \emph
  {et~al.}}]{Hou_2014}%
  \BibitemOpen
  \bibfield  {author} {\bibinfo {author} {\bibfnamefont {Z.}~\bibnamefont
  {Hou}} \emph {et~al.},\ }\href {\doibase 10.1088/0004-637x/782/2/74}
  {\bibfield  {journal} {\bibinfo  {journal} {The Astrophysical Journal}\
  }\textbf {\bibinfo {volume} {782}},\ \bibinfo {pages} {74} (\bibinfo {year}
  {2014})}\BibitemShut {NoStop}%
\bibitem [{\citenamefont {Hikage}\ \emph {et~al.}(2019)\citenamefont {Hikage}
  \emph {et~al.}}]{Hikage2019}%
  \BibitemOpen
  \bibfield  {author} {\bibinfo {author} {\bibfnamefont {C.}~\bibnamefont
  {Hikage}} \emph {et~al.},\ }\href {\doibase 10.1093/pasj/psz010} {\bibfield
  {journal} {\bibinfo  {journal} {Publications of the Astronomical Society of
  Japan}\ }\textbf {\bibinfo {volume} {71}} (\bibinfo {year} {2019}),\
  10.1093/pasj/psz010},\ \bibinfo {note} {43},\ \Eprint
  {http://arxiv.org/abs/https://academic.oup.com/pasj/article-pdf/71/2/43/28463847/psz010.pdf}
  {https://academic.oup.com/pasj/article-pdf/71/2/43/28463847/psz010.pdf}
  \BibitemShut {NoStop}%
\bibitem [{\citenamefont {{Akaike}}(1974)}]{AIC}%
  \BibitemOpen
  \bibfield  {author} {\bibinfo {author} {\bibfnamefont {H.}~\bibnamefont
  {{Akaike}}},\ }\href@noop {} {\bibfield  {journal} {\bibinfo  {journal} {IEEE
  Transactions on Automatic Control}\ }\textbf {\bibinfo {volume} {19}},\
  \bibinfo {pages} {716} (\bibinfo {year} {1974})}\BibitemShut {NoStop}%
\bibitem [{\citenamefont {Burnham}\ and\ \citenamefont
  {Anderson}(2004)}]{doi:10.1177/0049124104268644}%
  \BibitemOpen
  \bibfield  {author} {\bibinfo {author} {\bibfnamefont {K.~P.}\ \bibnamefont
  {Burnham}}\ and\ \bibinfo {author} {\bibfnamefont {D.~R.}\ \bibnamefont
  {Anderson}},\ }\href {\doibase 10.1177/0049124104268644} {\bibfield
  {journal} {\bibinfo  {journal} {Sociological Methods \& Research}\ }\textbf
  {\bibinfo {volume} {33}},\ \bibinfo {pages} {261} (\bibinfo {year} {2004})},\
  \Eprint {http://arxiv.org/abs/https://doi.org/10.1177/0049124104268644}
  {https://doi.org/10.1177/0049124104268644} \BibitemShut {NoStop}%
\end{thebibliography}%

\end{document}